\documentclass[prb,pre,aps,twocolumn,amsmath,amssymb,floatfix,
superscriptaddress]{revtex4-1}
\usepackage[dvips]{graphics}
\usepackage{graphicx}
\usepackage{color}
\usepackage{blindtext}
\usepackage{amsmath}
\definecolor{dred}{rgb}{0.75,0,0}
\usepackage{graphicx}
\usepackage{dcolumn}
\usepackage{bm}
\usepackage{xcolor}
\usepackage{braket}
\usepackage{physics}
\usepackage{appendix}
\usepackage{hyperref}
\hypersetup{
 colorlinks=true,
citecolor=magenta,filecolor=blue,
 linkcolor=magenta,
 }

\usepackage{graphicx}
\usepackage{dcolumn}
\usepackage{bm}
\usepackage{xcolor}

\definecolor{codegreen}{rgb}{0,0.6,0}
\definecolor{codegray}{rgb}{0.5,0.5,0.5}
\definecolor{codepurple}{rgb}{0.58,0,0.82}
\definecolor{backcolour}{rgb}{0.95,0.95,0.92}
 
\begin{document}

\preprint{APS/123-QED}

\title{\textcolor{dred}{Two-strand ladder network variants: localization, multifractality, and  quantum dynamics under an Aubry-Andr\'{e}-Harper kind of quasiperiodicity}} 

\author{Sougata Biswas}
\affiliation{Department of Physics, Presidency University, 86/1 College Street, Kolkata, West Bengal - 700 073, India}
\email{sougata.rs@presiuniv.ac.in}

\date{\today}

\begin{abstract}

In this paper we demonstrate, using a couple of variants of a two-strand ladder network that, a quasiperiodic 
 Aubry-Andr\'{e}-Harper (AAH)  modulation applied to the vertical strands, mimicking a deterministic distortion in the network, can give rise to certain exotic features in the electronic spectrum of such systems. While, for the simplest ladder network all the eigenstates become localized as the modulation strength crosses a threshold, for the second variant, modelling an ultrathin graphene nano-ribbon, the central part of the energy spectrum remains populated by extended wavefunctions. The multifractal character in the energy spectrum is observed for both these networks close to the critical values of the modulation. We substantiate our findings also by studying the quantum dynamics of a wave packet on such decorated lattices. Interestingly, while the mean square displacement (MSD) changes in the usual manner in a pure two-strand ladder network as the modulation strength varies, for the ultrathin graphene nanoribbon the temporal behaviour of the MSD remains unaltered only up to a strong modulation strength. This, we argue, is due to the extendedness of the wavefunction at the central part of the energy spectrum. Other measurements like the return probability, temporal autocorrelation function, the time dependence of the inverse participation ratio, and the information entropy are calculated for both networks with different modulation strengths and corroborate our analytical findings. 

\end{abstract}

\maketitle

\section{Introduction}
The investigation of disorder-driven wave localization, initiated by Anderson~\cite{anderson} and subsequently substantiated in numerous findings~\cite{lee,abrahams,borland}, has persistently attracted scholarly attention. Anderson observed that an electron loses phase coherence and the associated wavefunction generally becomes exponentially localized in three dimensions when there is an uncorrelated disorder in the distribution of potentials. However, there exists a threshold of disorder in three dimensions, below which the extended (metallic) character of the single-particle states can exist. Thus, a transition from a delocalized (metallic) phase to a localized (insulating) phase can be observed with a control over disorder. The term metal-insulator (MI) transition refers to this. The metal-insulator transitions are absent from one-dimensional Anderson models since all the single states are exponentially localized even under a nominal disorder~\cite{borland}. 

A metal-insulator transition is also observed in a one dimensional quasiperiodic model, viz, the Aubry-Andr\'{e}-Harper (AAH) model where, the quasiperiodicity enters through the potential landscape provided by the scatterers. However, the transition takes place in the parameter space~\cite{aubry,harper}, setting up a specific criterion between the values of the strength of the on-site potential and the nearest neighbor hopping integral, in the tight binding language. The eigenstates at the transition point are neither completely extended nor localized. They are termed {\it critical} in nature~\cite{ostlund}. Consequently, the system's behavior displays a broad range of distinct scaling exponents, that is, {\it multifractality}.

Research, based on AAH modulation and its variants has drawn wide attention over the years~\cite{danieli,an,devakul,roati}. The studies recently  included certain topology-related issues as well~\cite{madsen,dareau}. The dynamical properties of the generalized AAH model have also been extensively studied recently~\cite{purkayastha}.\par


\begin{figure}[ht]
\centering
\includegraphics[width=.9\columnwidth]{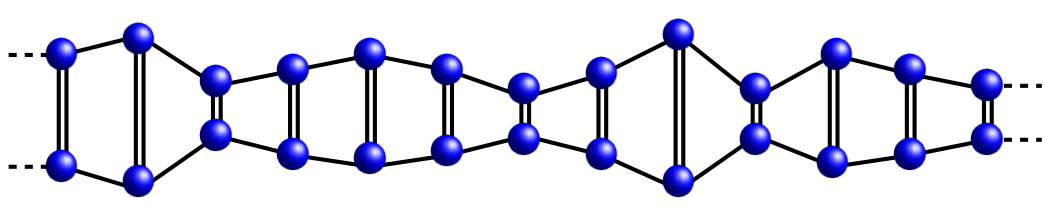}
\caption{(Color online) Schematic diagram of a distorted ladder network. }  
\label{ladder-dis}
\end{figure}

\begin{figure}[ht]
\centering
\includegraphics[width=.9\columnwidth]{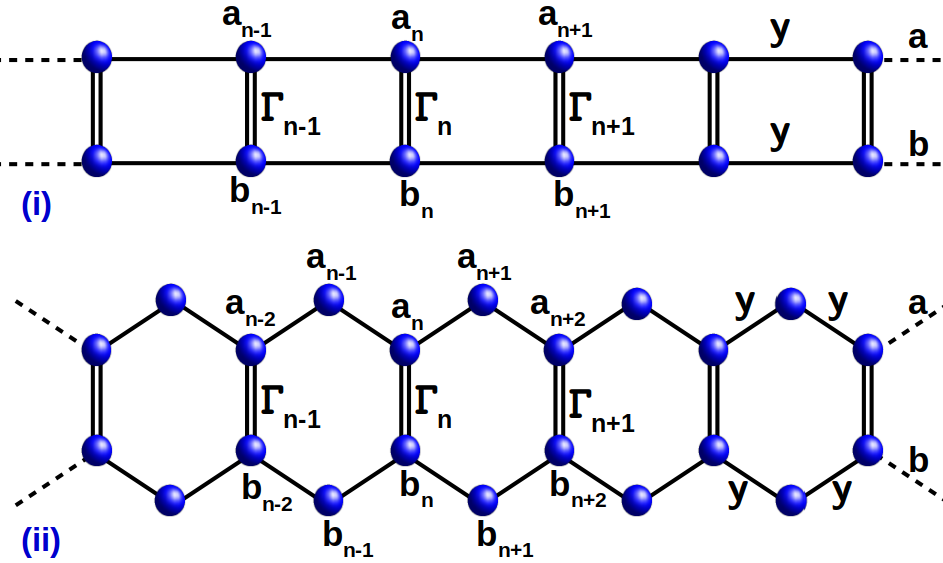}
\caption{(Color online) Schematic diagrams of (i) a simple two-strand ladder network, now straightened up for a better understanding, and (ii) an ultrathin graphene nano-ribbon (GNR) network }  
\label{figure}
\end{figure}

The metal-to-insulator transition due to an AAH kind modulation in the onsite potential term or even in the nearest neighbor hopping integrals is not new. However, the majority of the research is still confined to purely one-dimensional systems without any `width'. In the present communication, we intend to give a one-dimensional chain of atoms the most nominal width, which is easily done by designing a two-strand ladder model. In particular, we will be concerned with a {\it geometrically distorted ladder} (GDL). In the tight binding language, such a distortion is easily achieved by assigning different values of the hopping amplitude along the `rungs' of the ladder. In this paper, such `distortion' will be deterministic, and will be mimicked by an AAH type of modulation. The results so far, to the best of our knowledge, are not known, at least in good detail and this aspect serves as the central motivation behind the present work. Will there be any metal-to-insulator transition under such AAH modulation in the vertical hopping part of such low-dimensional decorated lattices? We intend to scrutinize and find an answer to this question. 

We first address the question of localization or delocalization or a so-called `distortion-driven' metal-insulator transition for a simple two-strand ladder network. For a completely random distortion, such a network can be schematically represented by Fig~\ref{ladder-dis}. Later, we will draw it as shown in Fig.~\ref{figure} (i), just for the sake of an easy understanding of the system. 

We continue our investigation on a second kind of ladder network which will be an approximation to say, an ultrathin graphene nanoribbon (GNR).  We observe that, while all the eigenstates of the ladder network in the first case become localized after a particular modulation strength, interestingly, the central region of the eigenspectrum of the GNR always remains populated by delocalized (extended) eigenfunctions, even if the modulation strength is large enough. As a result, a complete metal-to-insulator transition is not possible in an ultrathin GNR network under such AAH modulation. Additionally, we have explored that both the distorted networks exhibit multifractality in the neighbourhood of their critical (transition) points.

Secondly, we examine the role of the strength of the AAH modulation in the quantum dynamics of a wave packet released on such two-strand ladders. The study includes the spreading of wavepacket with time, long-time behaviour of the mean square displacement (MSD), the return probability, temporal autocorrelation function, information entropy, and inverse participation ratio. To the best of our knowledge, these dynamical aspects are practically unaddressed for such quasi-one-dimensional ladder geometries under an AAH modulation.

Our results on quantum dynamics indicate that, for the pure states, exhibited by the basis two-strand ladder network, the time exponent of MSD varies from $2$ to $0$ as the modulation strength increases, and the corresponding return probability changes in the usual manner. However, the wavepacket dynamics are completely different for an ultrathin GNR network. This is a result of having a spectrum comprising both localized and extended eigenstates.  We will discuss that, for the mixed states the de-localized part of the wavepacket controls the MSD, whereas the localized part of the wavepacket plays an important role in controlling other quantities like return probability, inverse participation ratio, or temporal autocorrelation function. Therefore, in the particular case of an ultrathin GNR although the MSD indicates that the system is under ballistics motion, at the same time it has a finite return probability. This is a result that hasn't been encountered before. 

Fig.~\ref{figure} contains two lattice models, (i) ladder network and (ii) ultrathin graphene nano-ribbon (GNR) network.  The behaviour of the eigenstates of the simple ladder network~\cite{sil,santanu,sil1} and ultrathin GNR network~\cite{brey,katsunori} are already available in the literature. In this paper, as we have given the Aubry-André-Harper (AAH) kind of modulation profile in the hopping along the rungs, the networks become very useful in low-dimensional physics for modeling DNA-like biological systems~\cite{arunava} and may provide answers to several issues~\cite{klotsa,wang,rakitin,pablo}. 

The plan of the paper is as follows. In Section II, the model system and tight binding  Hamiltonian are presented. In Section III, we will discuss the localization-delocalization features of the eigenstates with the help of the inverse participation ratio and investigate the multifractal nature at the transition (critical) point. In section IV, we mainly focus on the quantum dynamics of the wavepacket by calculating the spreading of the wavepacket, mean square displacement (MSD), return probability, temporal auto-correlation function, information entropy, and time evolution of inverse participation ratio. Finally, we draw our conclusion in section V.

\section{The model and Hamiltonian}
We refer to Fig.~\ref{figure}, where a basic two-strand ladder network (in (i)), and an ultrathin graphene nanoribbon (in (ii)) are shown. Two arms are denoted by $a$ and $b$. At first, we incorporate two types of hopping in these models. One is the nearest neighbour hopping $y$ between the $n^{th}$ and $(n+1)^{th}$ sites of every arm of the ladder and ultrathin GNR networks along $x$ direction and another is the vertical hopping $\Gamma$ between the $n^{th}$ sites of the two arms $a$ and $b$. 
The tight binding Hamiltonian for spinless, non-interacting fermions in Wannier basis can be written as, 
\begin{equation} 
H = y \sum_{n} (a_{n+1}^{\dag} a_{n} + b_{n+1}^{\dag} b_{n}) +\epsilon \sum_{n}(a_{n}^{\dag} a_{n} + b_{n}^{\dag} b_{n}) + \Delta
\label{hamiltonian}
\end{equation}
Where $ \Delta = \Gamma \sum_{n}(a_{n}^{\dag} b_{n} + b_{n}^{\dag} a_{n})$ and this part of the hamiltonian is different for these two networks. For the simple ladder network, this summation index $n$ goes as $n, n+1, n+2,....$, whereas it is $n, n+2, n+4,....$ for ultrathin GNR network. \\  
Here, $\epsilon$ symbolizes the on-site potential and is set to zero in all our numerical calculations. We have given the Aubry-Andr\'{e}-Harper modulation profile in each vertical (rung) hopping $\Gamma_{n} =  y (1+\lambda \cos~(2 \pi Q n))$ (with $n = 1, 2, 3,...$). $\lambda$ is the strength of the modulation and $Q = \frac{\sqrt{5}-1}{2}$ is an irrational number (incommensurate limit). We choose the lattice constant as unity.\\


\begin{figure}[ht!]
\centering
(a)\includegraphics[width=.44\columnwidth]{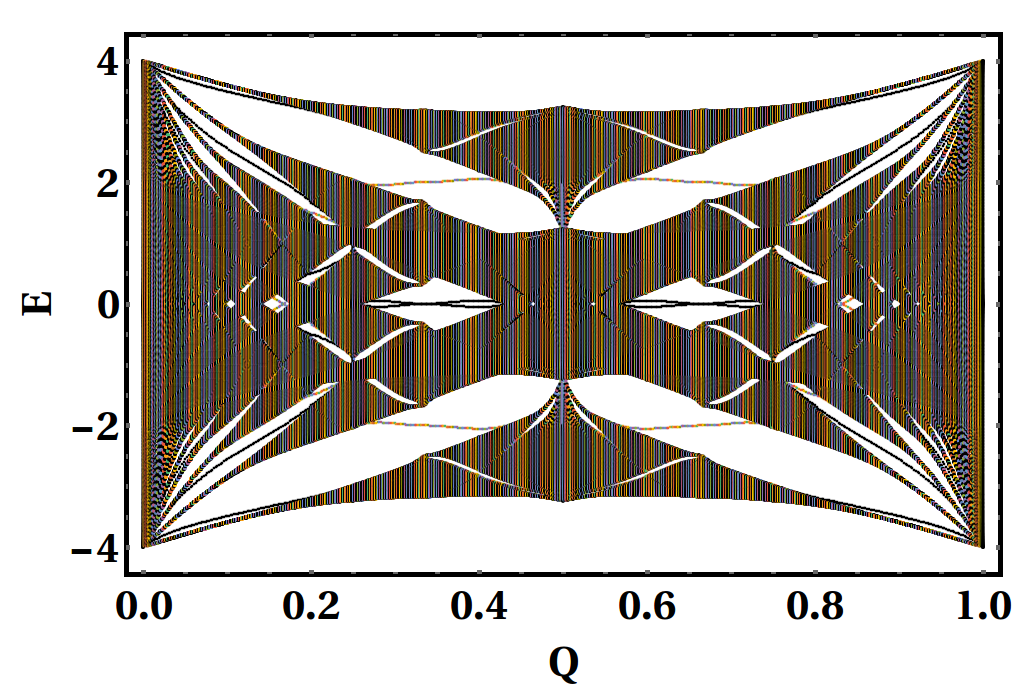}
(b)\includegraphics[width=.44\columnwidth]{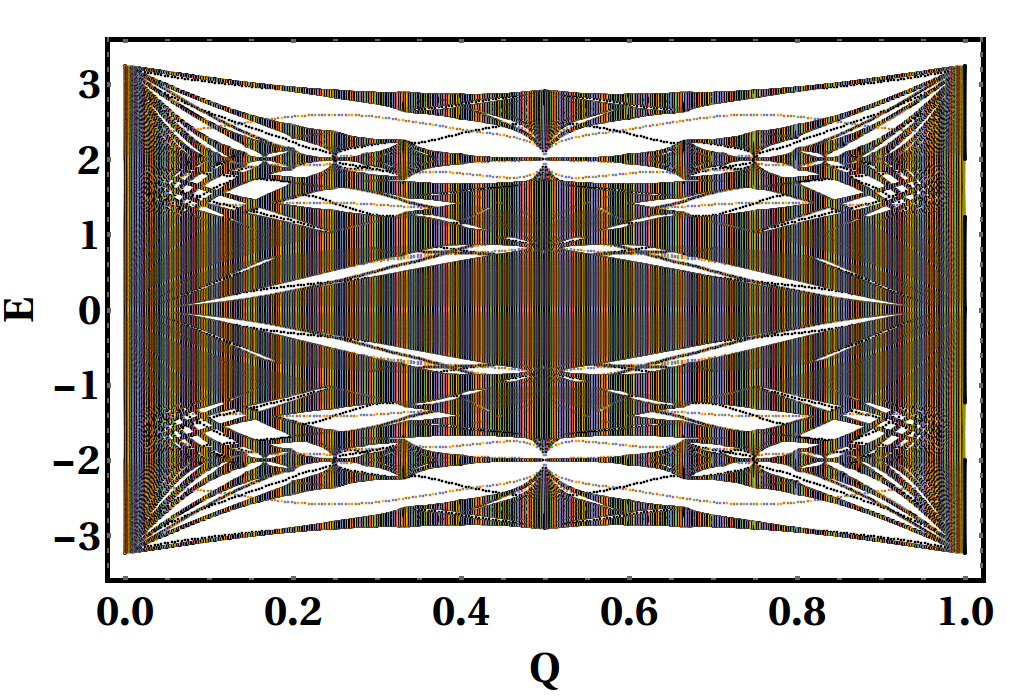}

\caption{(Color online) Energy spectra as a function of $Q$ for (a) ladder network ($1002$ sites), (b) ultrathin graphere nano-ribbon ($1002$ sites). The rung hopping parameter is given by   $\Gamma_{n}= y (1+ \lambda \cos~( 2 \pi Q n))$ with $\lambda = 1$.  The onsite potential $\epsilon$ is chosen as zero. The parameter $ y $ is set as $1$. }  
\label{e-Q}
\end{figure}
\begin{figure}[ht!]
\centering
(a)\includegraphics[width=.44\columnwidth]{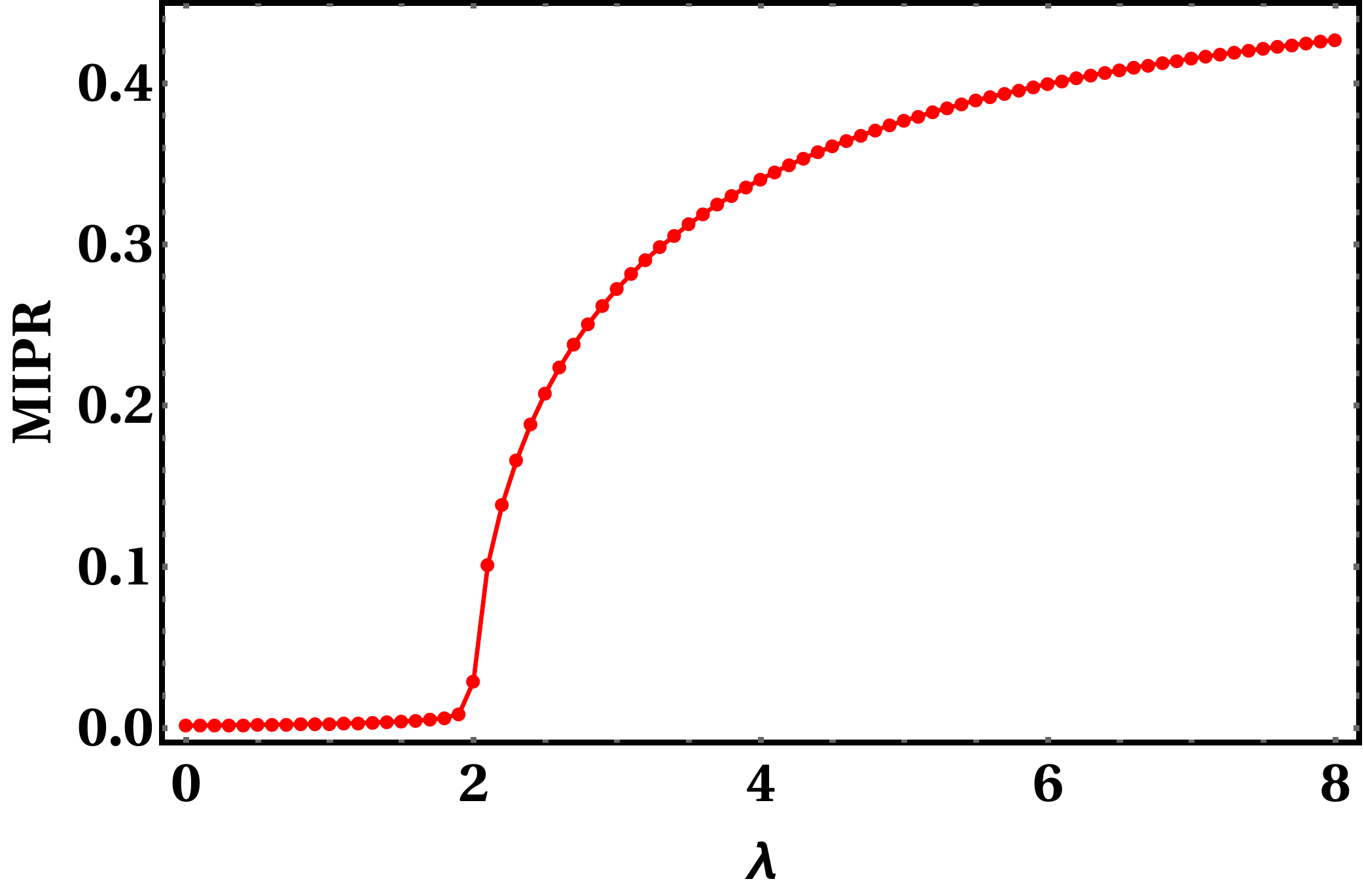}
(b)\includegraphics[width=.44\columnwidth]{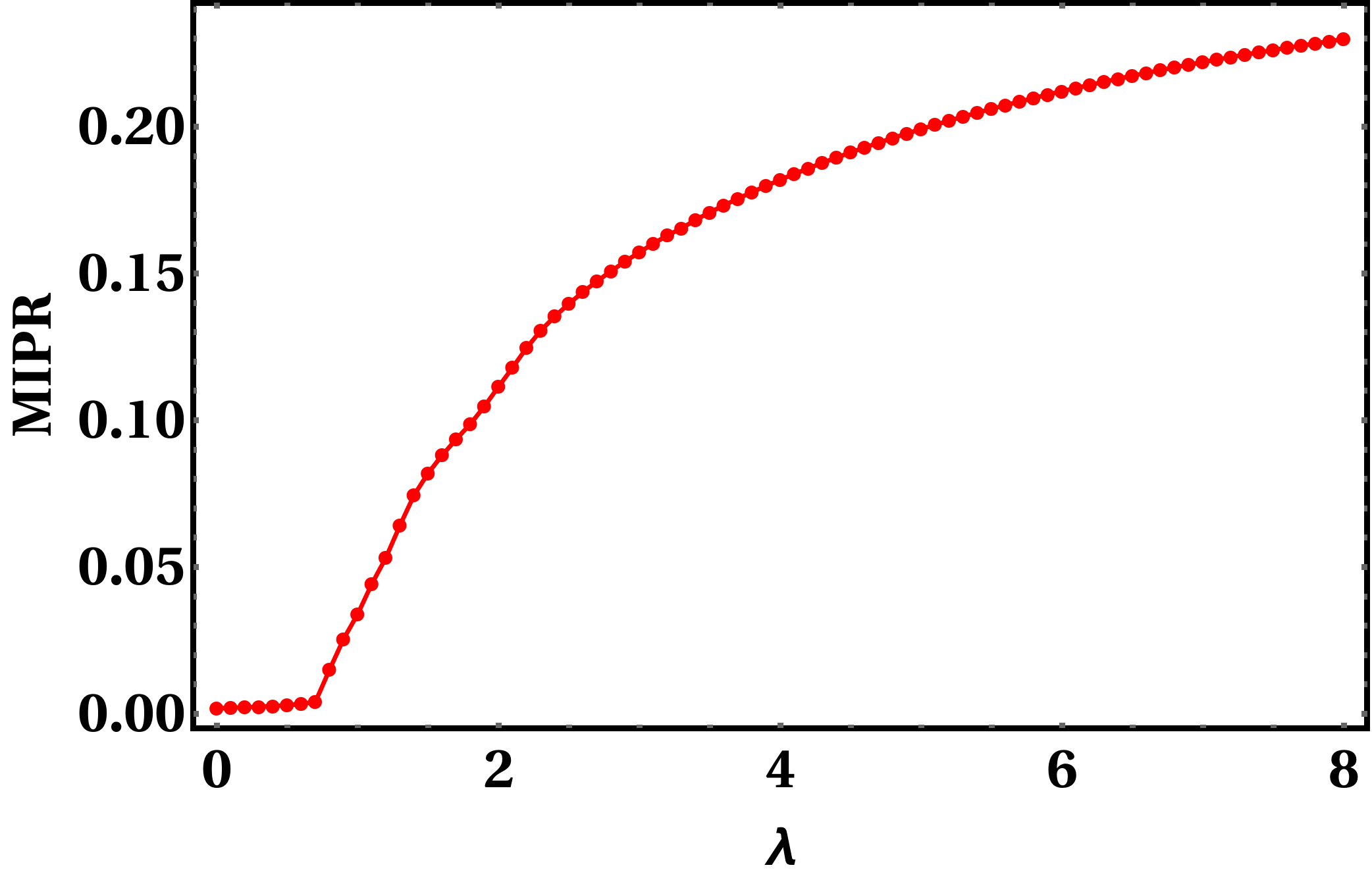}

(c)\includegraphics[width=.44\columnwidth]{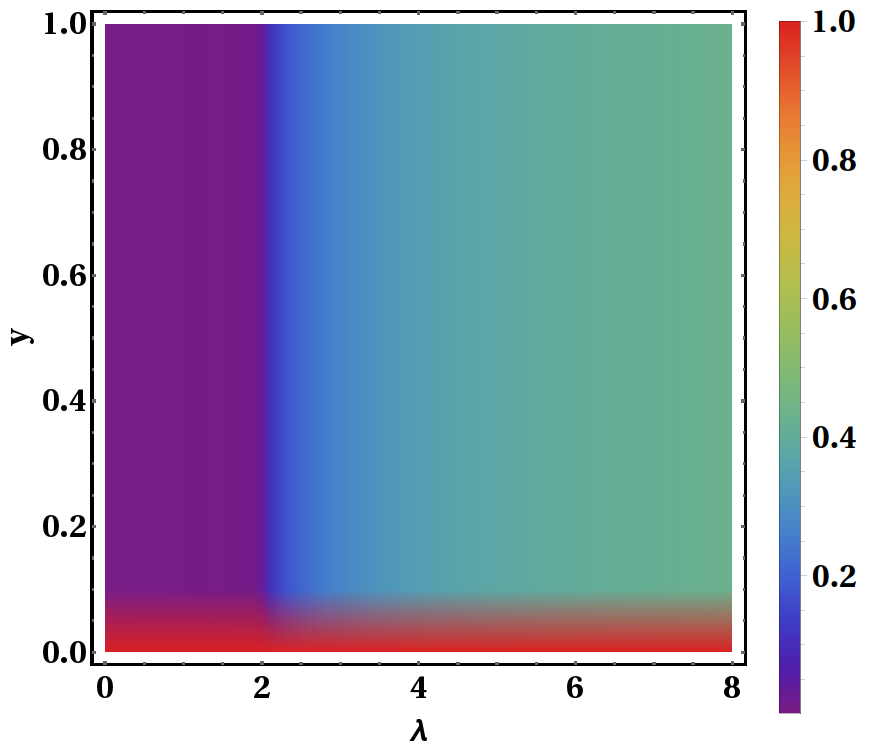}
(c)\includegraphics[width=.44\columnwidth]{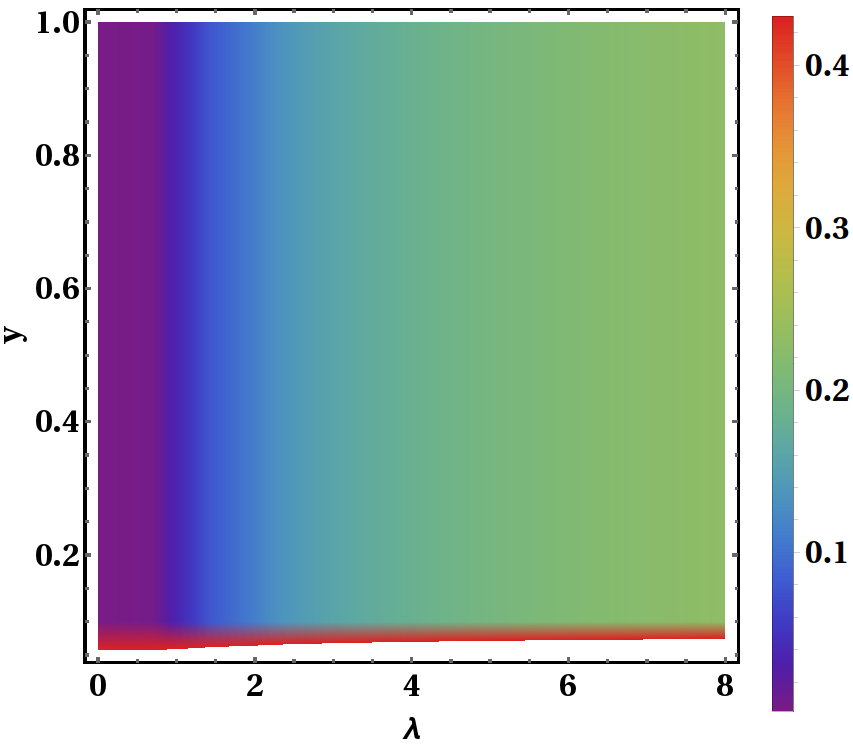}

(e)\includegraphics[width=.44\columnwidth]{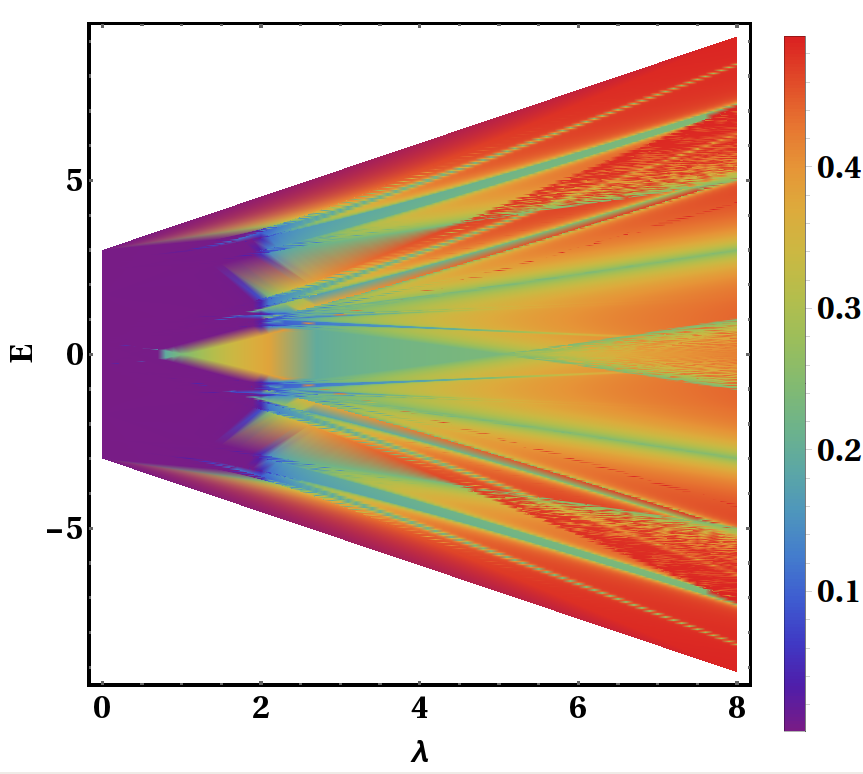}
(f)\includegraphics[width=.44\columnwidth]{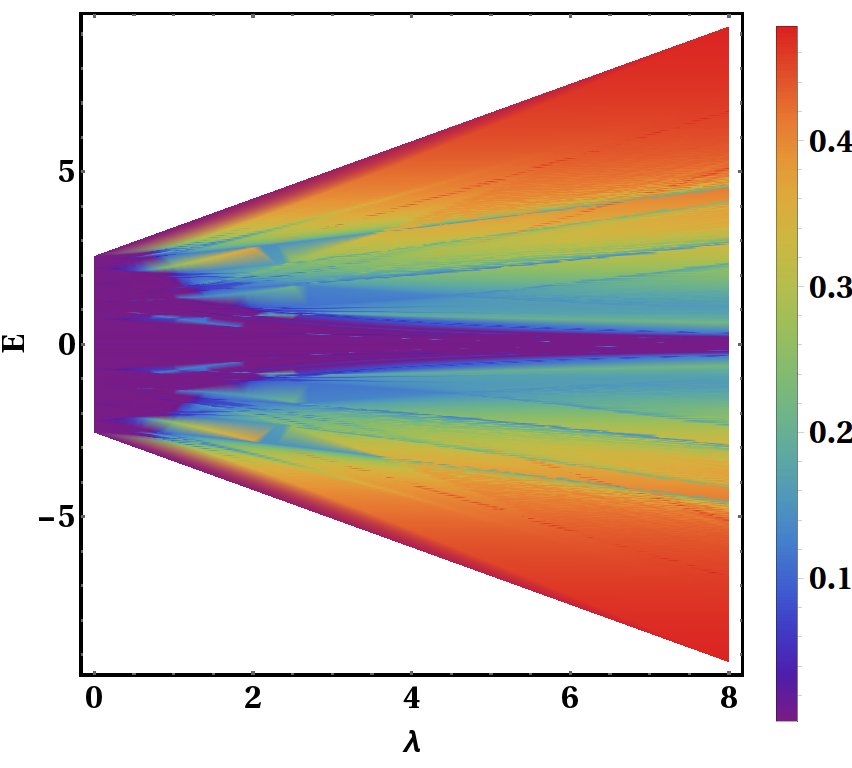}

\caption{(Color online) Mean Inverse Participation Ratio (MIPR) is plotted against $\lambda$ for (a) ladder network ($1002$ sites), (b) ultrathin graphene nano-ribbon network ($1002$ sites). The rung hopping parameter is given by $\Gamma_{n}= y (1+ \lambda \cos~( 2 \pi Q n))$, where $Q = \frac{\sqrt{5}-1}{2}$. The onsite potential $\epsilon$ is chosen as zero. The parameter $ y $ is set as $1$. (c,d) Density plot of MIPR within the parameter space $y$ and $\lambda$ for ladder and ultrathin GNR network respectively. (e,f) Density plot of IPR with energy ($E$) and $\lambda$ for ladder and ultrathin GNR network respectively.}  
\label{mipr}
\end{figure}


\section{Localization and De-localization of the eigenstates}
In this section, we will characterize de-localized, localized, and multifractal regions of the energy spectrum of both ladder and ultrathin GNR networks depending upon the strength of the Aubry modulation $\lambda$. Before going into detail, we have tried to discuss the energy spectrum of these quantum networks as the function of Q. Fig.~\ref{e-Q}  shows the distribution of energy eigenvalues as a function of $Q$ with an interval $0.001$ for the ladder and ultrathin GNR network and got the famous Hofstadter butterfly structure~\cite{Hofstader}. 

\subsection{Mean Inverse Participation Ratio (MIPR)}
The inverse participation ratio (IPR) is a popular measurement for detecting the localization properties of the eigenstates. It is defined as the fourth power of the normalized wave function~\cite{tong,izrailev}.
\begin{equation}
    IPR = \sum_n |\psi_n|^4
    \label{IPR-eq}
\end{equation}
where $n$ runs over all atomic sites. It gives us an understanding of the distribution of the amplitudes of the wave function over atomic sites. For example, for a completely de-localized eigenstate, the IPR goes to zero, whereas it has a finite value (tends to unity), for localized eigenstates. The mean or average IPR ($MIPR$) is defined as, $\frac{1}{N}\sum_{n=1}^{N}IPR_{n}$, where $N$ is the total number of atomic sites~\cite{Joana,Qi}. \par
The variation of MIPR with the modulation strength $\lambda$ is depicted in  Fig.~\ref{mipr}(a, b) for ladder and ultrathin GNR quantum networks respectively. It is seen that the MIPR is zero in a particular range of $\lambda$, which is known as a de-localized region. Further increment of $\lambda$, MIPR is no longer zero, rather it contains a small finite value after a particular value of $\lambda$. This is called transition (critical) point $\lambda_c$.  The corresponding density plot of MIPR against the hopping term $y$ and modulation strength $\lambda$ is shown in Fig.~\ref{mipr}(c,d). From these plots, it is clear that a transition from de-localized to localized eigenstates occurs depending upon the modulation strength, and the transition point is located at $\lambda_c \approx 2$, and $ \approx 0.75$ for the ladder, and ultrathin GNR network respectively.\par 
Fig.~\ref{mipr}(e,f) represents the density plot of IPR against each energy eigenvalue with different strengths of AAH modulation $\lambda$. The Fig.~\ref{mipr}(e) corresponds to the ladder model, which clearly demonstrates that all IPR practically goes to zero when $\lambda<\lambda_c$ and all IPR contain finite values at $\lambda > \lambda_c$. This special type of IPR distribution of all energy eigenvalues with the modulation strength $\lambda$ ensures the occurrence of metal-to-insular phase transition in parameter space. Fig.~\ref{mipr}(f) represents similar plots for the ultrathin GNR network. Again for $\lambda<\lambda_c$ all IPR are zero. But when  $\lambda > \lambda_c$ IPR have a special character. The IPR corresponding to the central energy part is still zero whereas the IPR takes finite values in two sides of the central energy region. As a result, the central energy region always exhibits a de-localized character. So a pure metal-to-insulator transition is not possible in such an ultrathin GNR network.

\begin{figure}[ht!]
\centering
\includegraphics[width=.8\columnwidth]{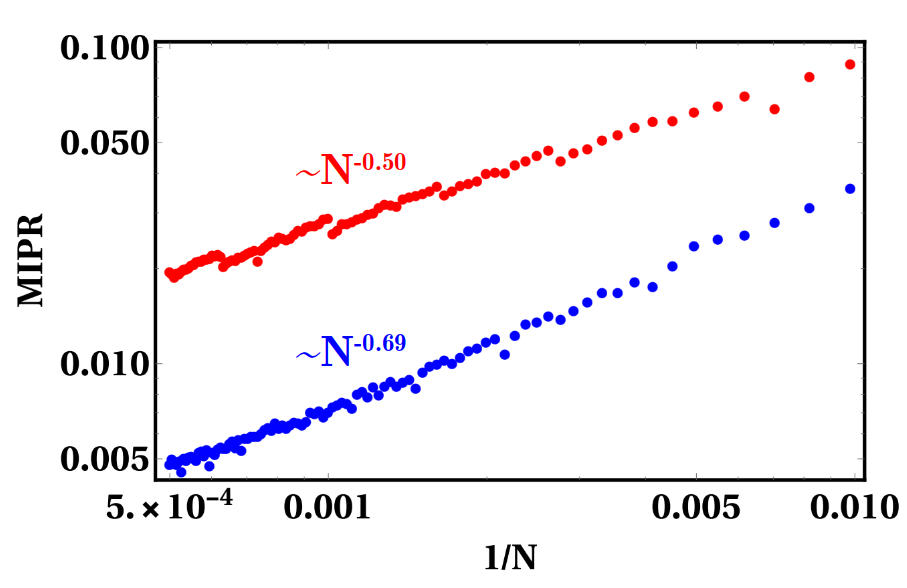}

\caption{(Color online) Finite-size scaling of the mean IPR for ladder network (red colored) and ultrathin graphene nano-ribbon (blue colored) in their critical region i.e, $\lambda = 2$ for ladder network and $\lambda = 0.75 $ for ultrathin GNR network. The rung hopping parameter is given by $\Gamma_{n}= y (1+ \lambda \cos~( 2 \pi Q n))$, where $Q = \frac{\sqrt{5}-1}{2}$. The onsite potential $\epsilon$ is chosen as zero. The parameter $ y $ is set as $1$.   }  
\label{scaling}
\end{figure}



\begin{figure}[ht!]
\centering
(a)\includegraphics[width=.44\columnwidth]{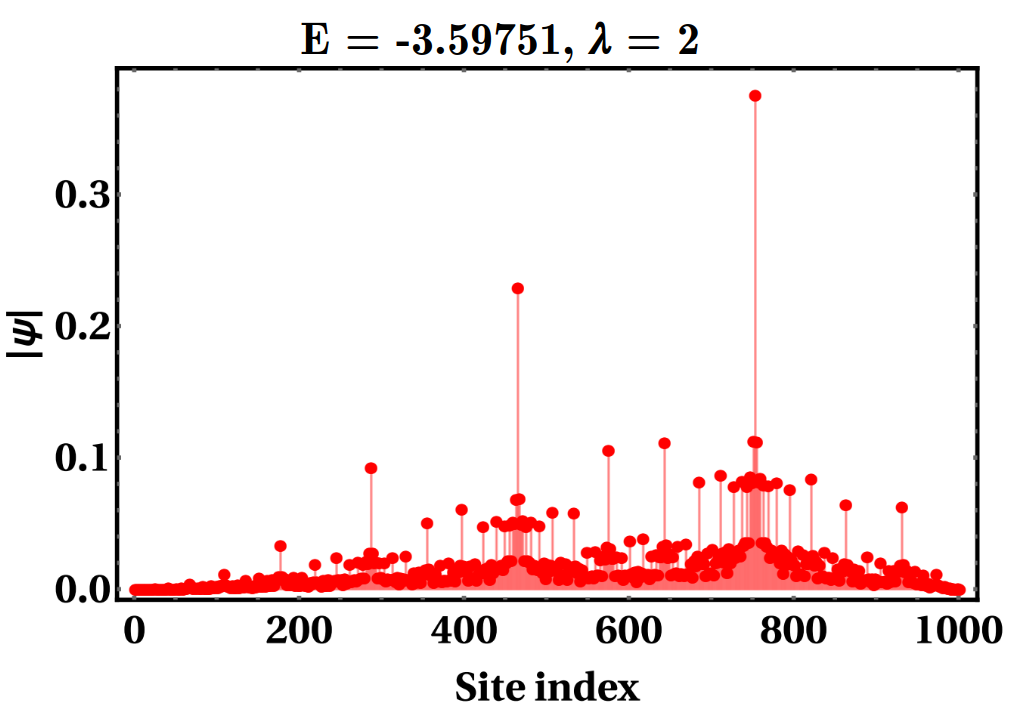}
(b)\includegraphics[width=.44\columnwidth]{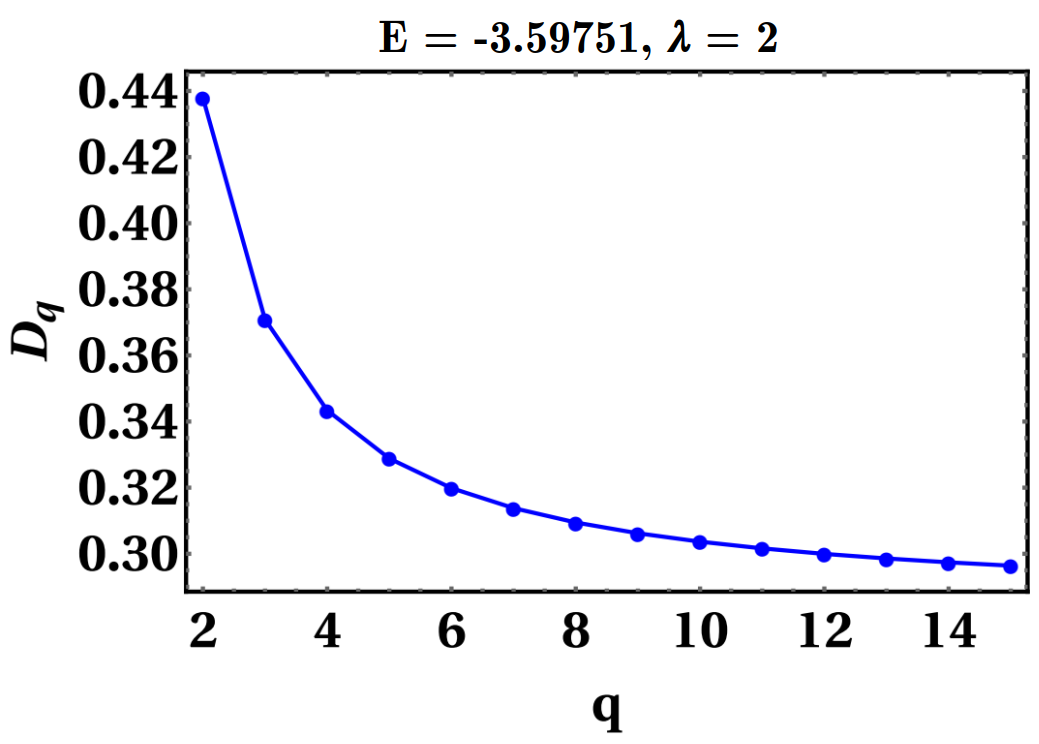}
(c)\includegraphics[width=.44\columnwidth]{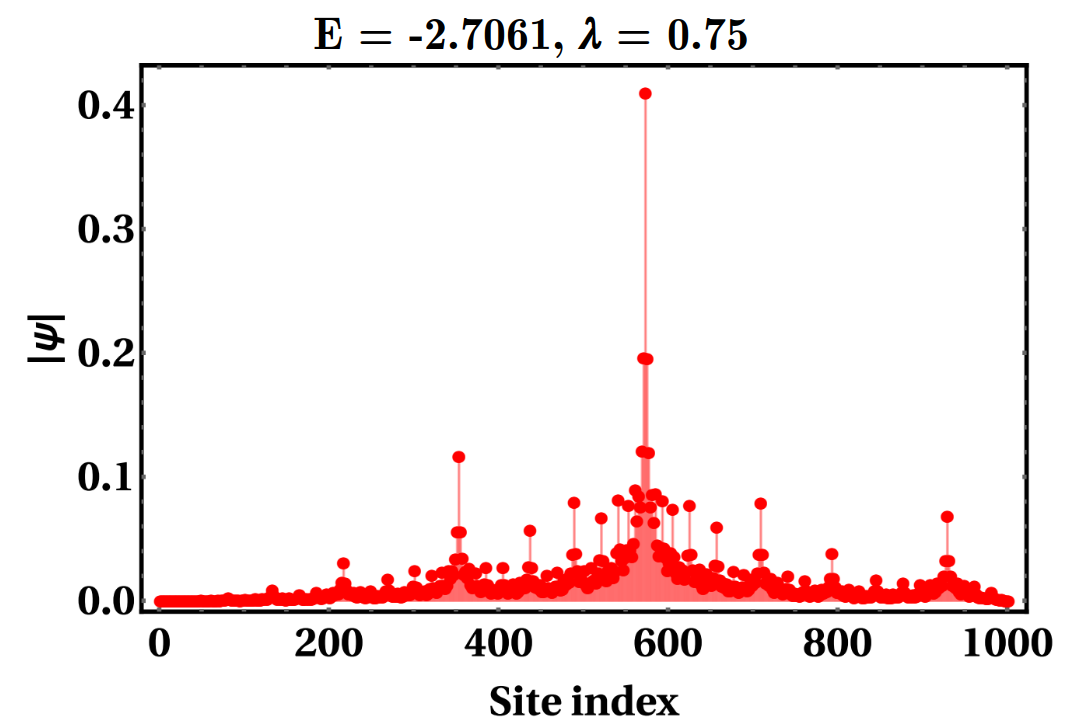}
(d)\includegraphics[width=.44\columnwidth]{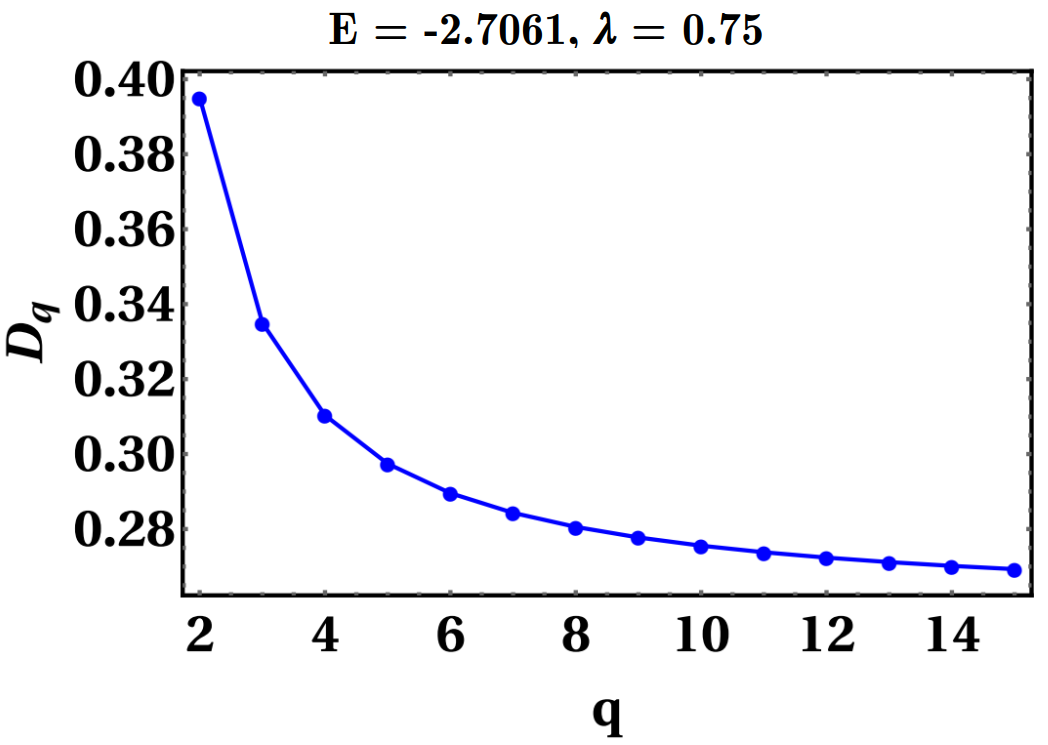}

\caption{(Color online) (a,c) Distribution of the ground state energy wavefunction amplitude for (a) ladder network ($1002$ sites), (c) ultrathin graphene nano-ribbon ($1002$ sites) within the critical region. (b,d) are the corresponding $D_q$.vs. $q$ plots. The rung hopping parameter is given by $\Gamma_{n}= y (1+ \lambda \cos~( 2 \pi Q n))$, where $Q = \frac{\sqrt{5}-1}{2}$. The onsite potential $\epsilon$ is chosen as zero. The parameter $ y $ is set as $1$. The strength $\lambda$ is set as $2$, and $0.75$ for (a,b), and (c,d) respectively.}  
\label{multi}
\end{figure}



\begin{figure}[ht]
\centering
(a)\includegraphics[width=.44\columnwidth]{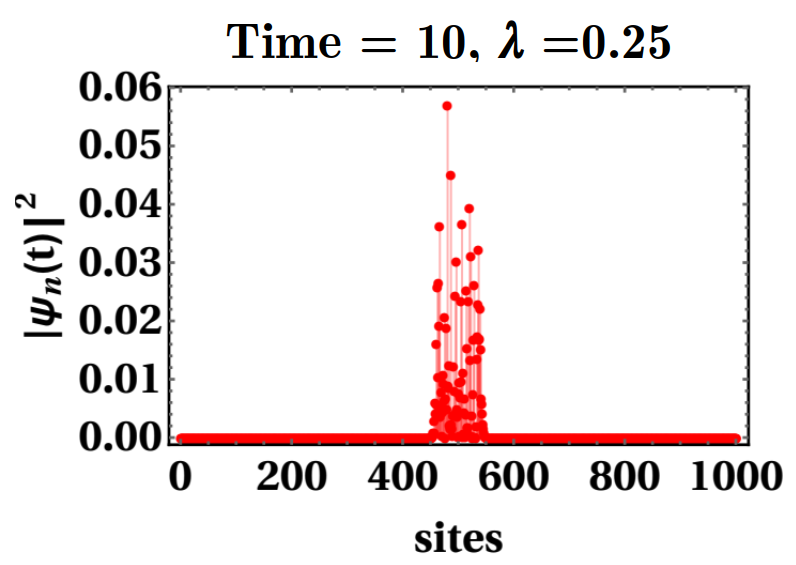}
(b)\includegraphics[width=.44\columnwidth]{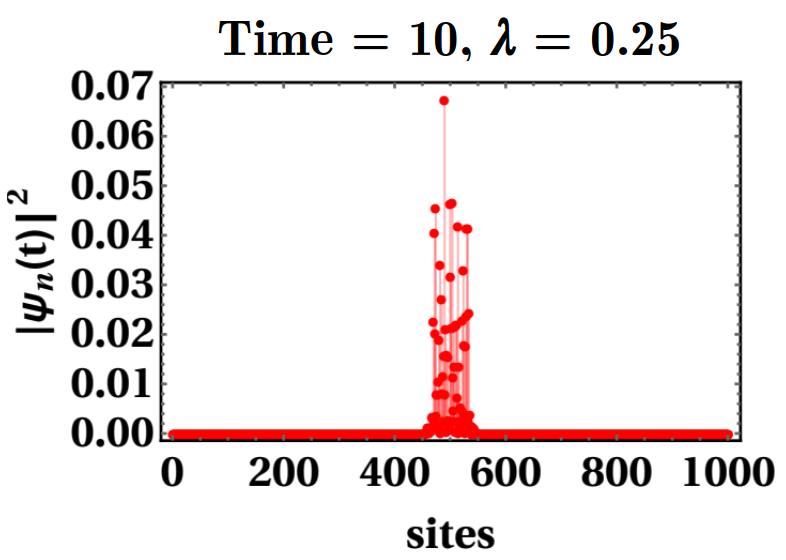}
(c)\includegraphics[width=.44\columnwidth]{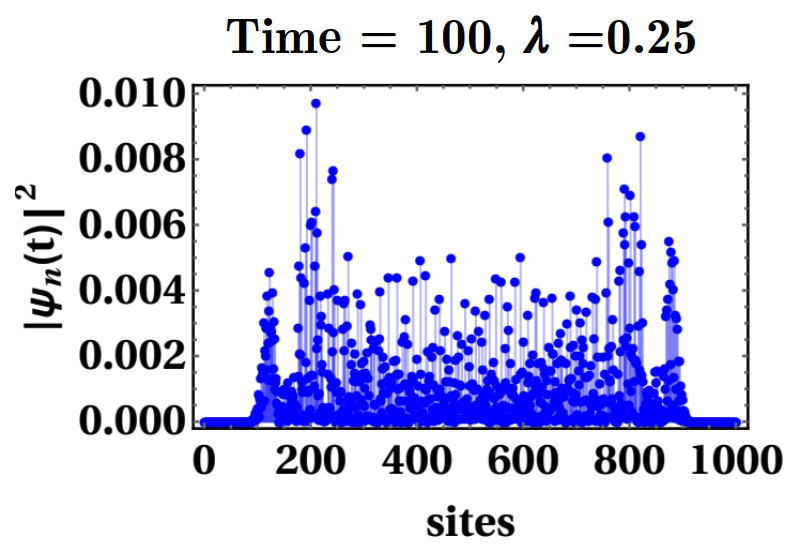}
(d)\includegraphics[width=.44\columnwidth]{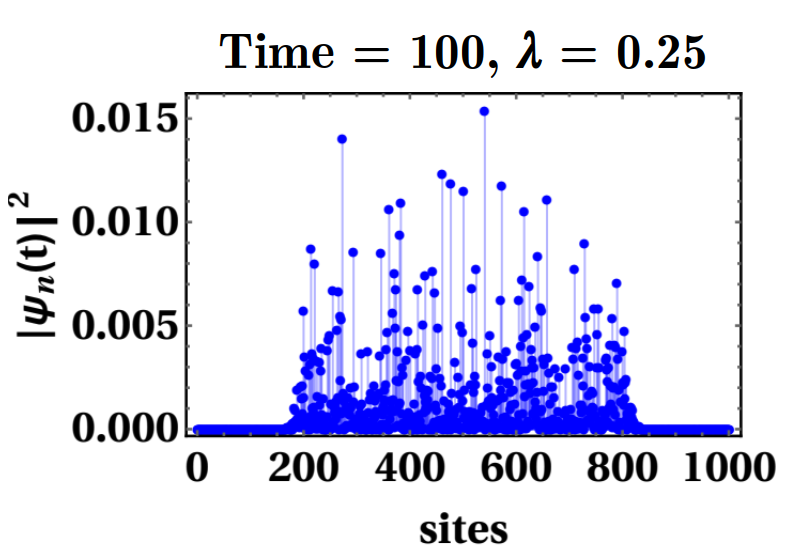}
(e)\includegraphics[width=.44\columnwidth]{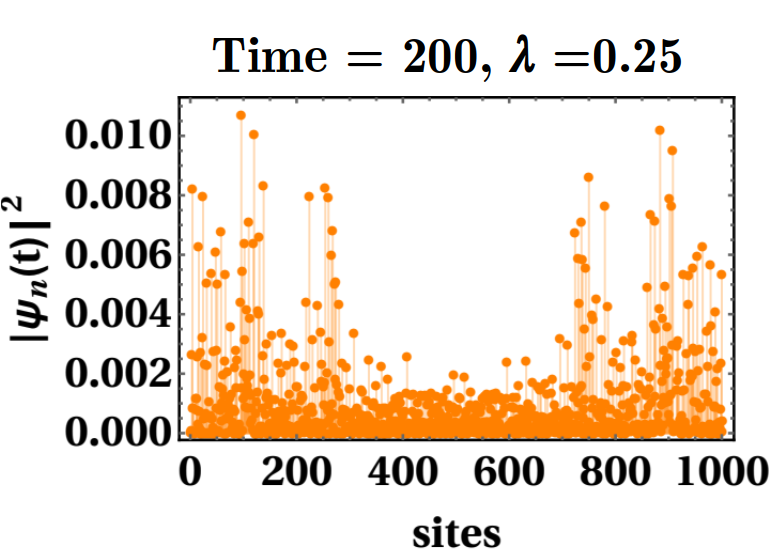}
(f)\includegraphics[width=.44\columnwidth]{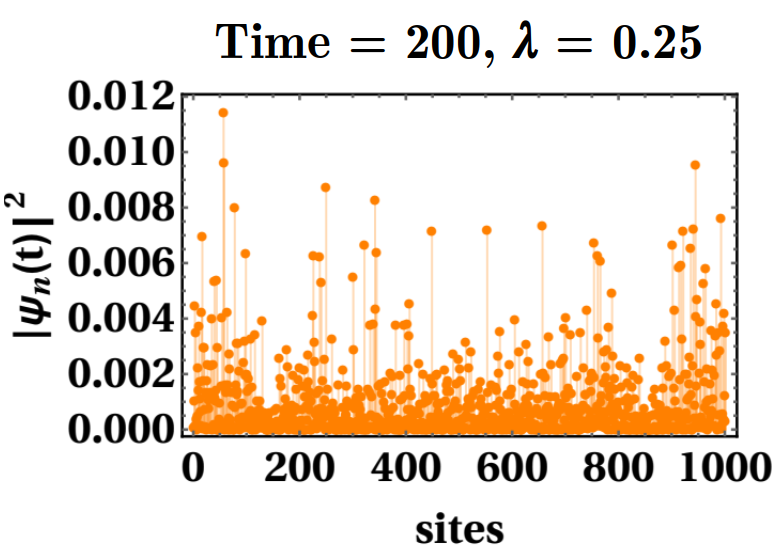}

\caption{(Color online) The variation of the spatial distribution of the wave packet for (a, c ,e) ladder network ($1002 sites$), and (b, d, f) ultrathin graphene nano-ribbon ($1002 sites$)  with $time = 10$(red colored), $time = 100$(blue colored), and $time = 200$(orange colored) respectively. Initially, we put the wave packet on the $501^{th}$ site. The rung hopping parameter is given by $\Gamma_{n}= y (1+ \lambda \cos~( 2 \pi Q n))$, where $Q = \frac{\sqrt{5}-1}{2}$. The strength of the modulation is chosen as $\lambda = 0.25$. The onsite potential $\epsilon$ is chosen as zero. The parameter $ y $ is set as $1$. }
\label{psin1}
\end{figure}


\begin{figure*}[ht!]
\centering
(a)\includegraphics[width=.44\columnwidth]{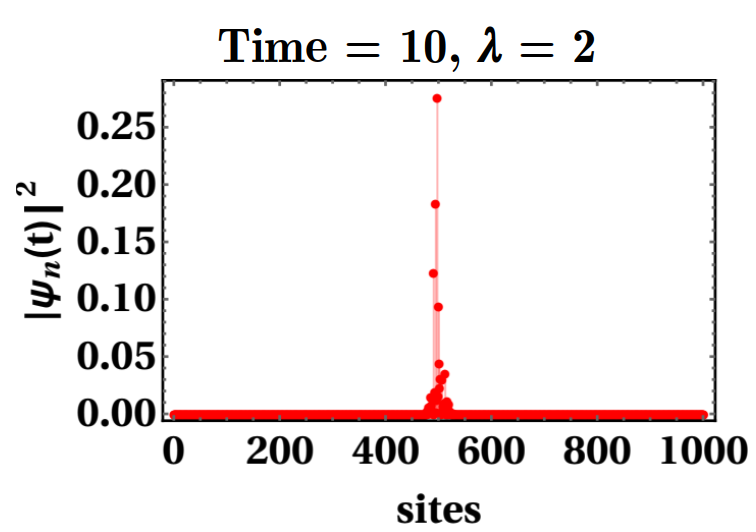}
(b)\includegraphics[width=.44\columnwidth]{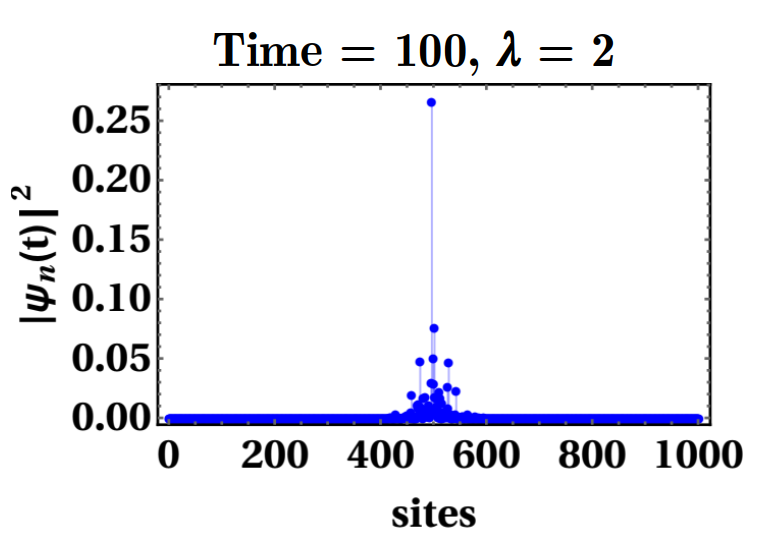}
(c)\includegraphics[width=.44\columnwidth]{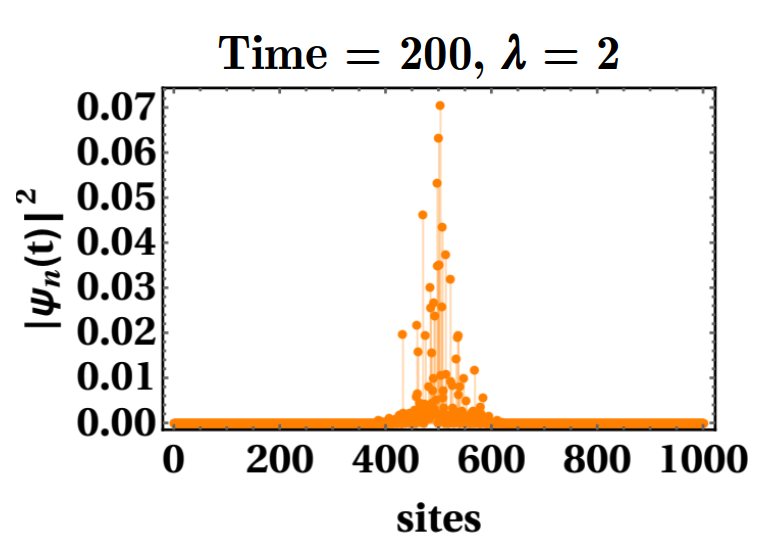}
(d)\includegraphics[width=.44\columnwidth]{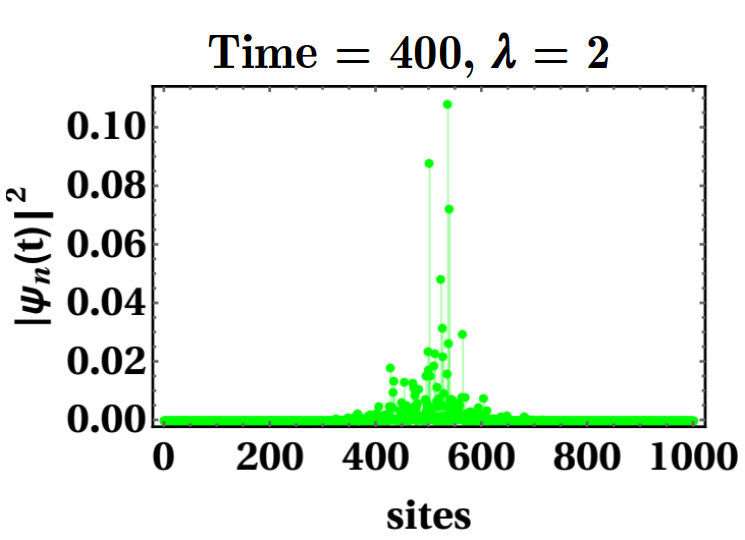}
(e)\includegraphics[width=.44\columnwidth]{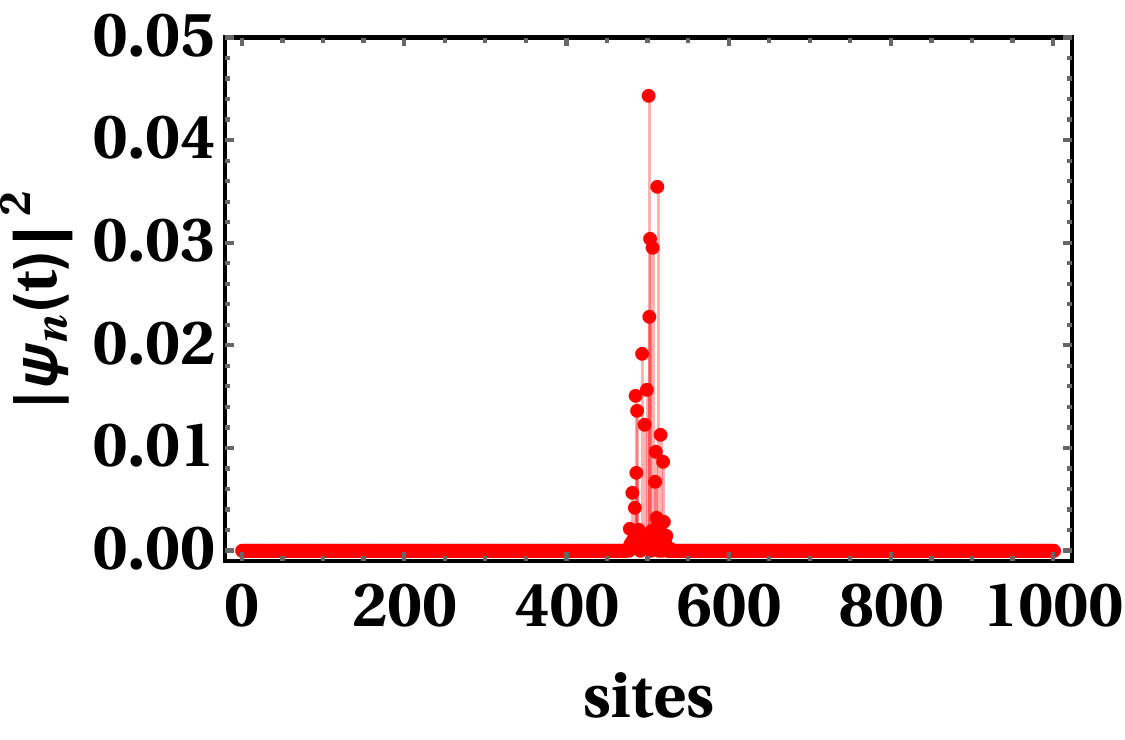}
(f)\includegraphics[width=.44\columnwidth]{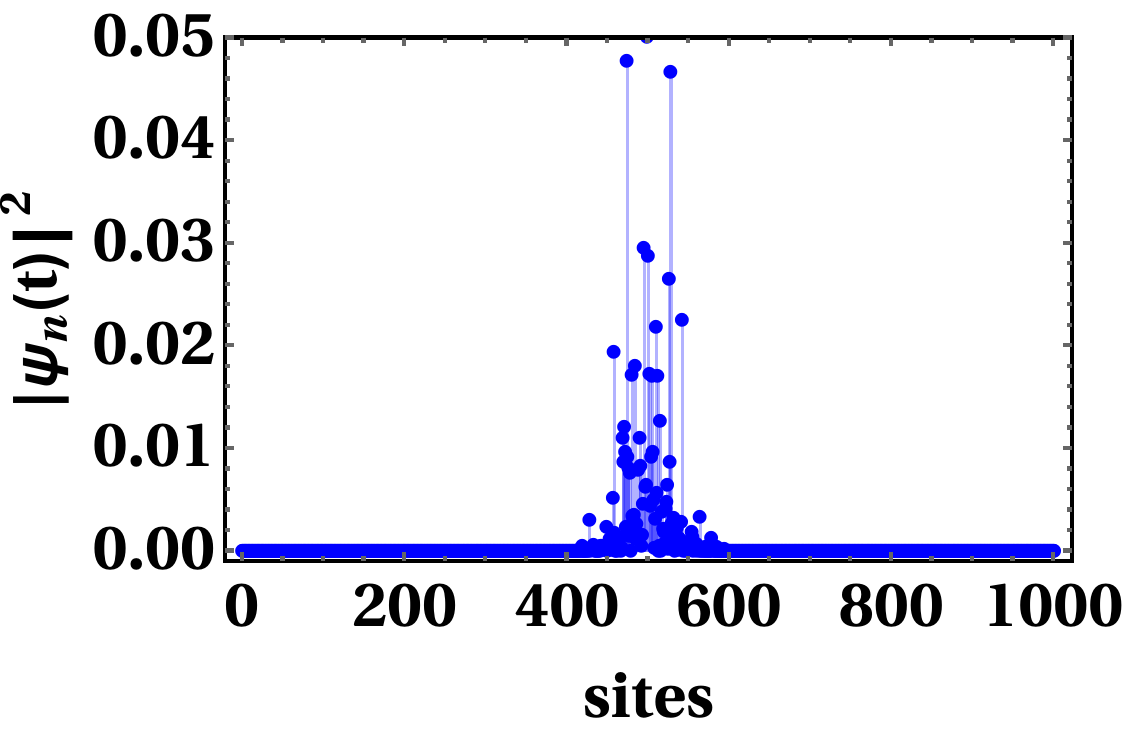}
(g)\includegraphics[width=.44\columnwidth]{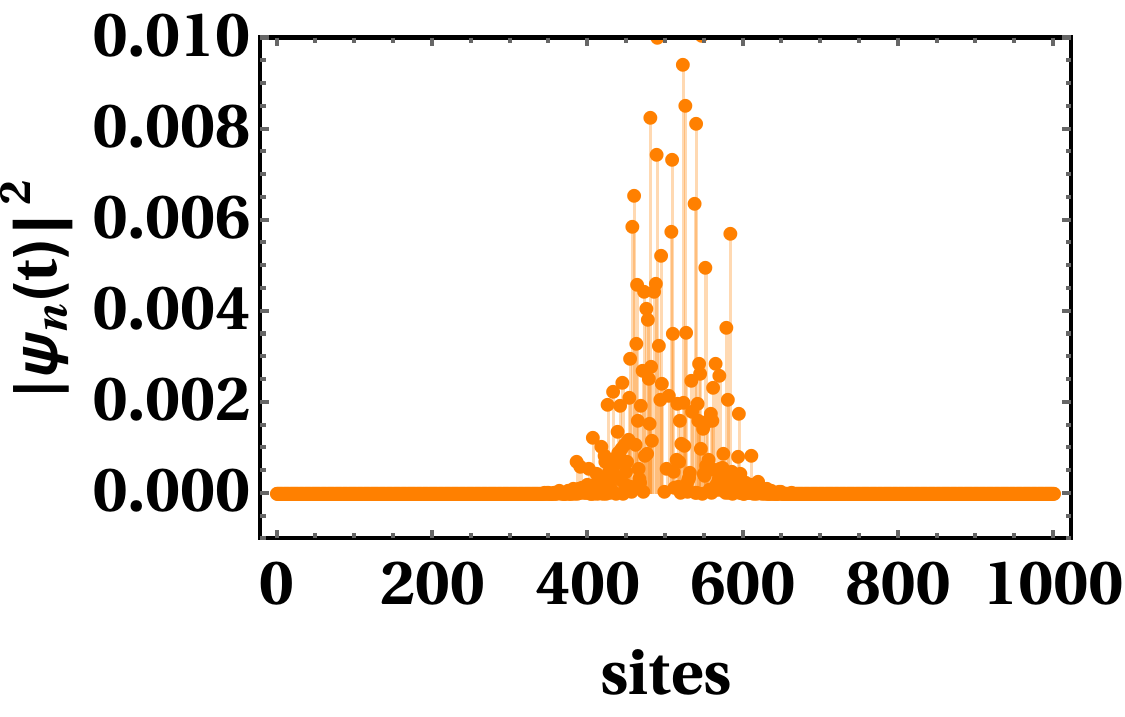}
(h)\includegraphics[width=.44\columnwidth]{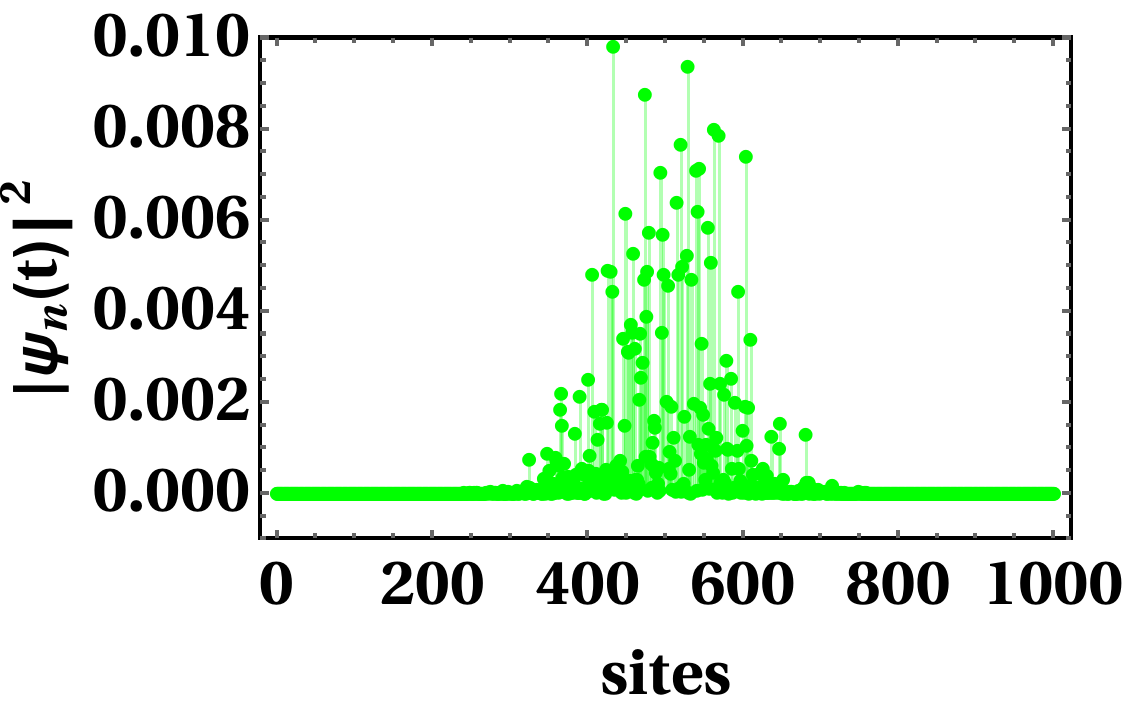}
(i)\includegraphics[width=.44\columnwidth]{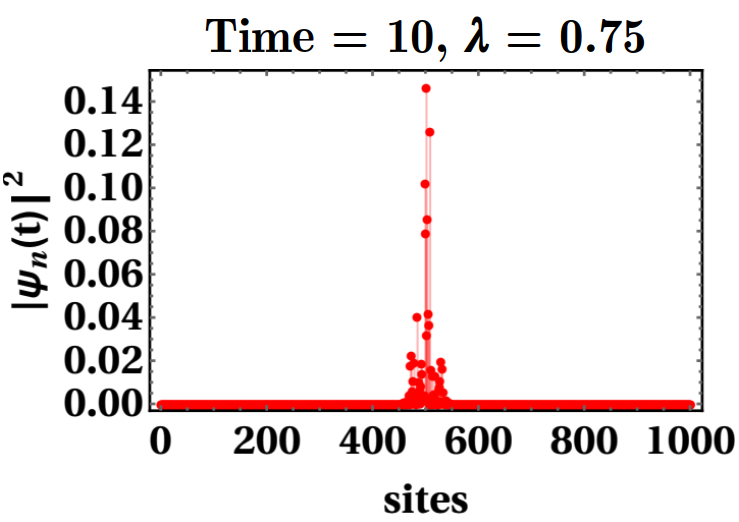}
(j)\includegraphics[width=.44\columnwidth]{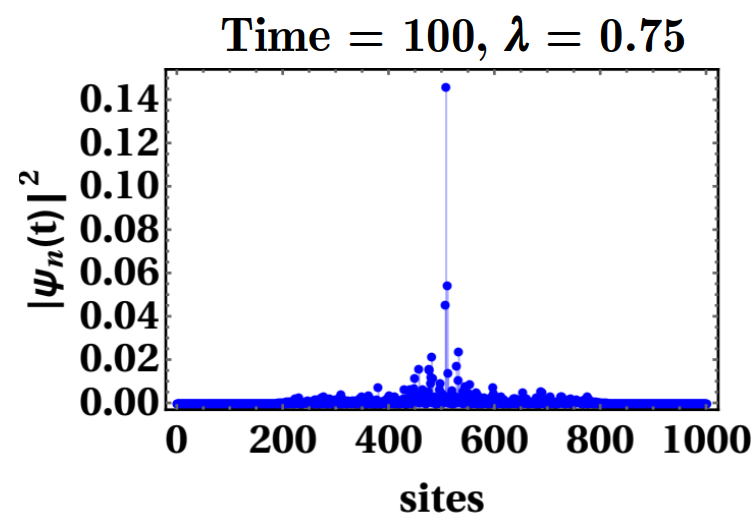}
(k)\includegraphics[width=.44\columnwidth]{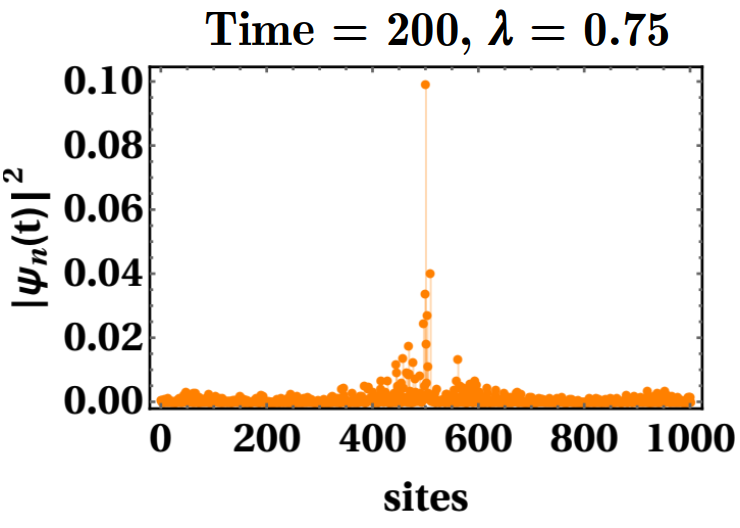}
(l)\includegraphics[width=.44\columnwidth]{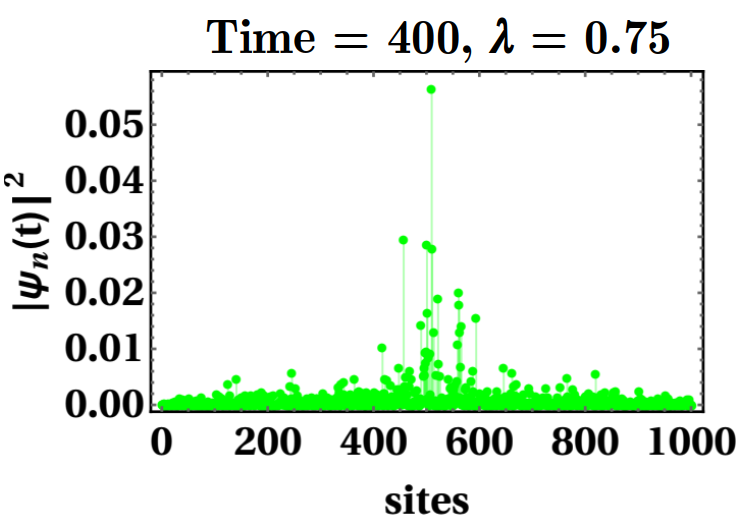}
(m)\includegraphics[width=.44\columnwidth]{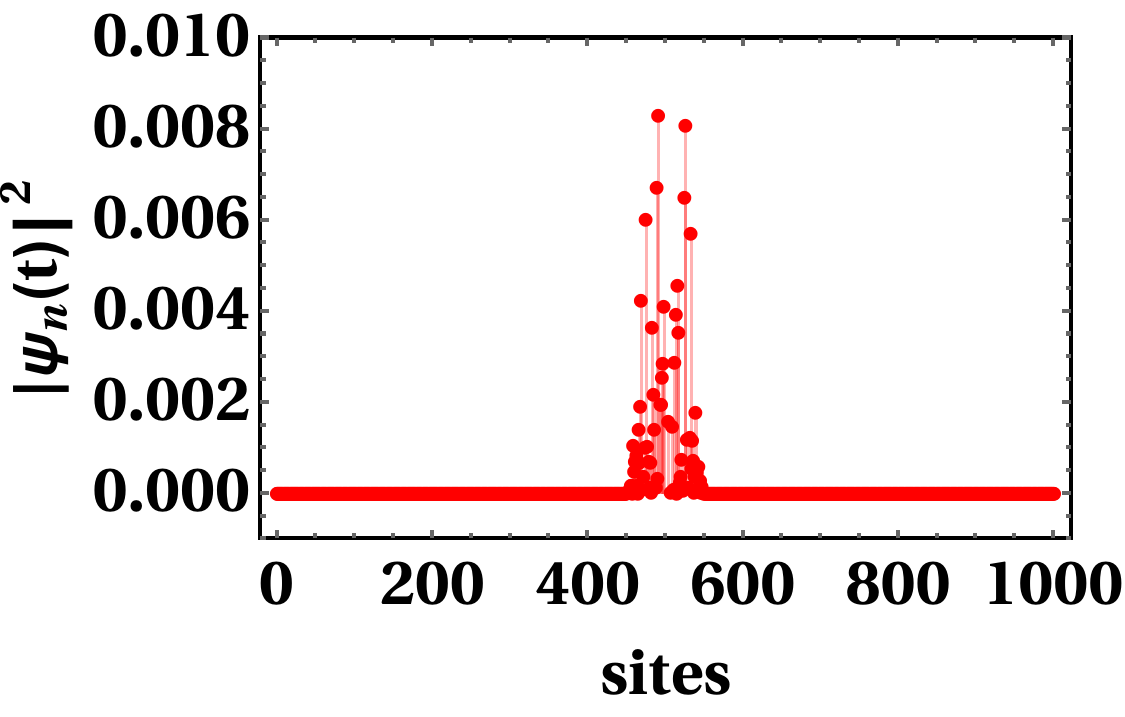}
(n)\includegraphics[width=.44\columnwidth]{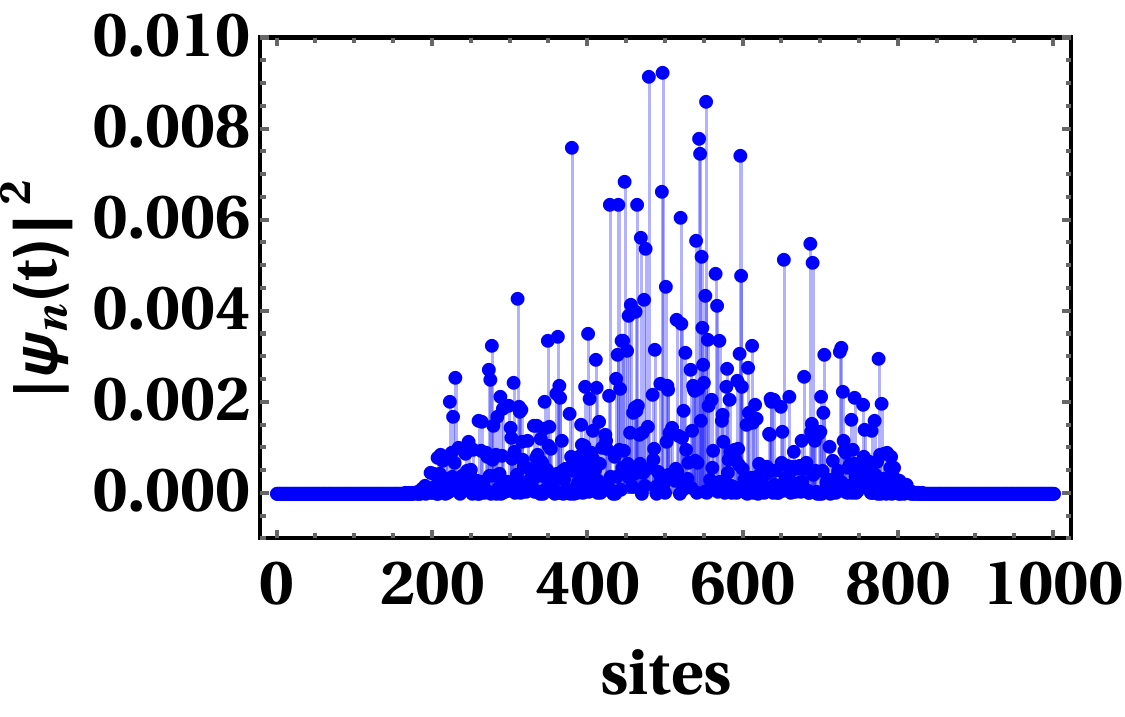}
(o)\includegraphics[width=.44\columnwidth]{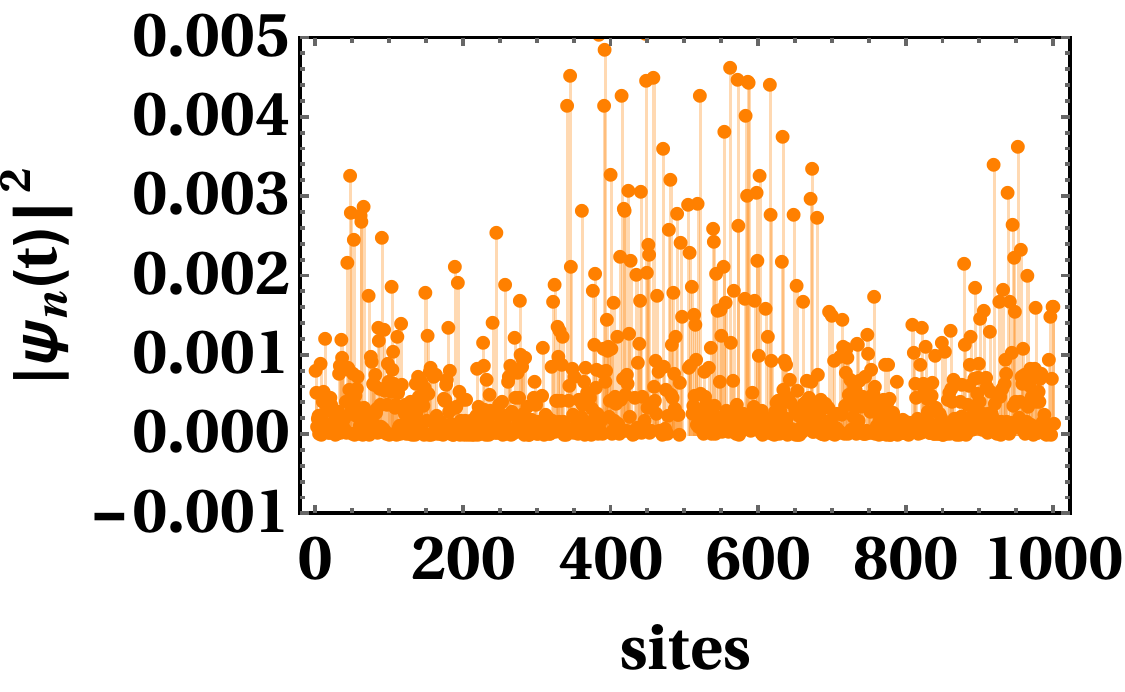}
(p)\includegraphics[width=.44\columnwidth]{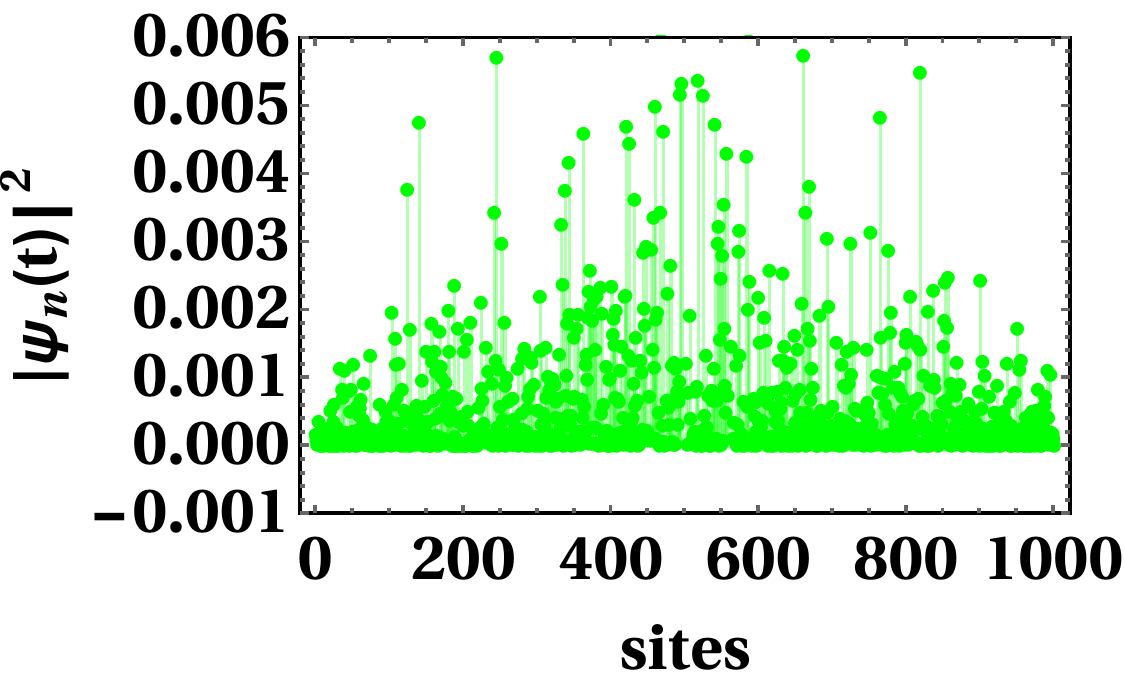}
\caption{(Color online) The variation of the spatial distribution of the wave packet for (a-h) ladder network ($1002 sites$), and (i-p) ultrathin graphene nano-ribbon network ($1002 sites$)  with $time = 10$(red colored), $time = 100$(blue colored), and $time = 200$(orange colored) and $time = 400$(green colored) respectively. Initially, we put the wave packet on the $501^{th}$ site. The rung hopping parameter is given by $\Gamma_{n}= y (1+ \lambda \cos~( 2 \pi Q n))$, where $Q = \frac{\sqrt{5}-1}{2}$. The strength of the modulation is chosen as $\lambda = 2$ for the ladder network and $\lambda = 0.75$ for the graphene nanoribbon network. The onsite potential $\epsilon$ is chosen as zero. The parameter $ y $ is set as $1$. (e,f,g,h) and (m,n,o,p) are the magnified versions of (a,b,c,d) and (i,j,k,l) respectively.}
\label{psin2}
\end{figure*}

\begin{figure*}[ht!]
\centering
(a)\includegraphics[width=.44\columnwidth]{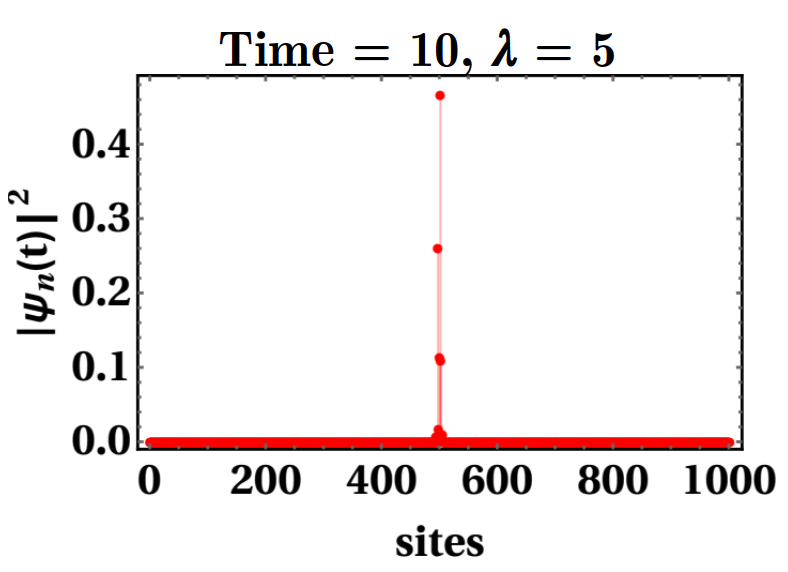}
(b)\includegraphics[width=.44\columnwidth]{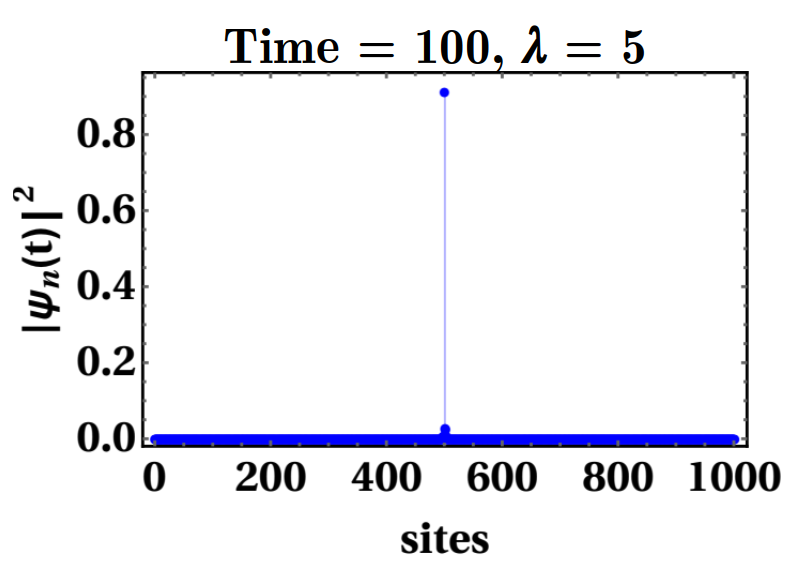}
(c)\includegraphics[width=.44\columnwidth]{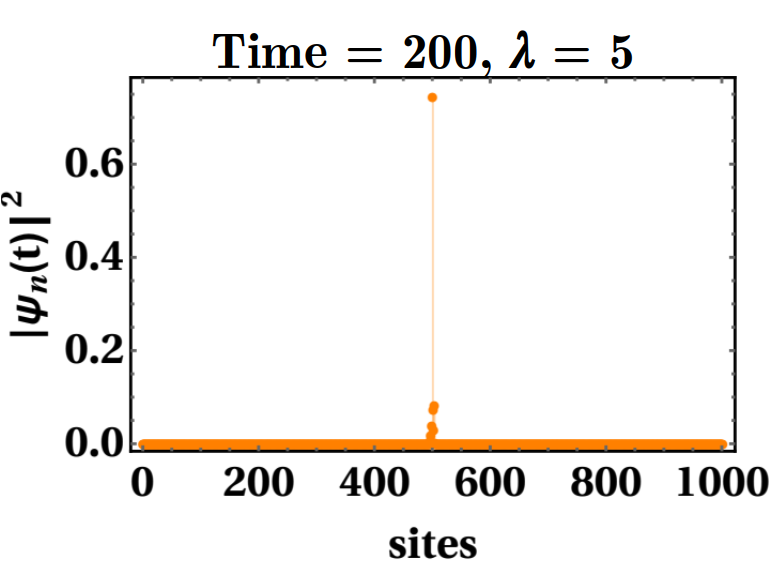}
(d)\includegraphics[width=.44\columnwidth]{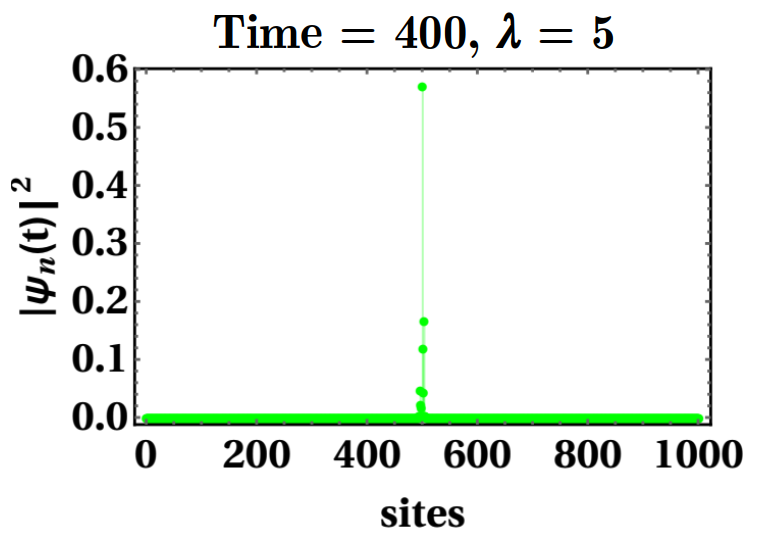}
(e)\includegraphics[width=.44\columnwidth]{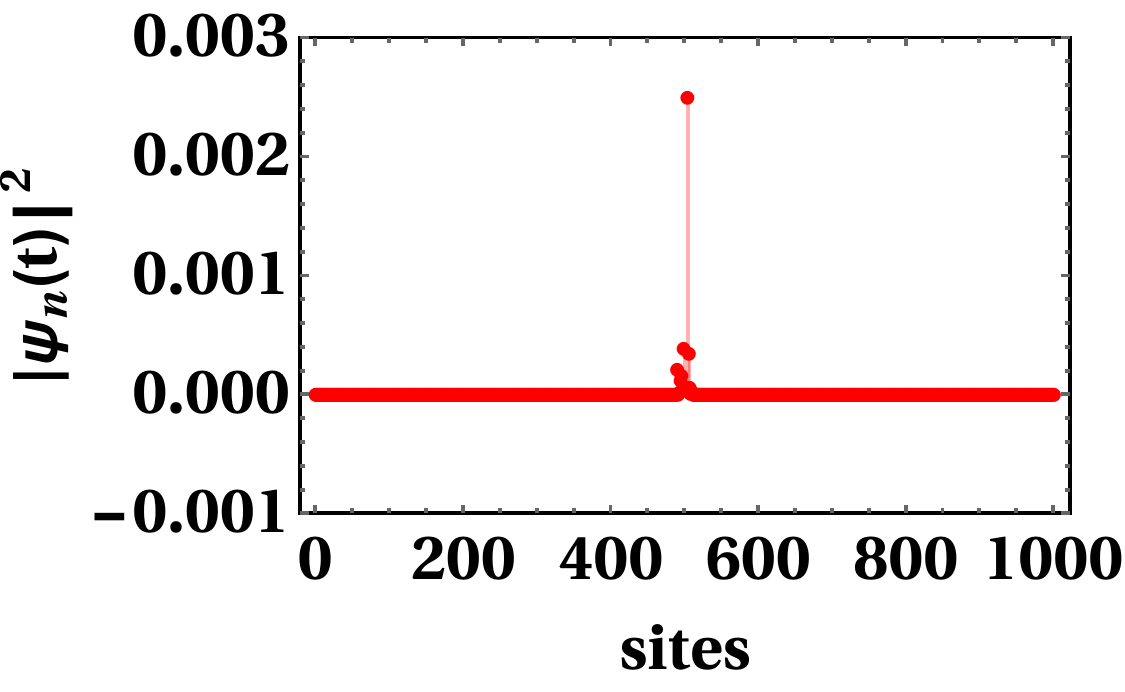}
(f)\includegraphics[width=.44\columnwidth]{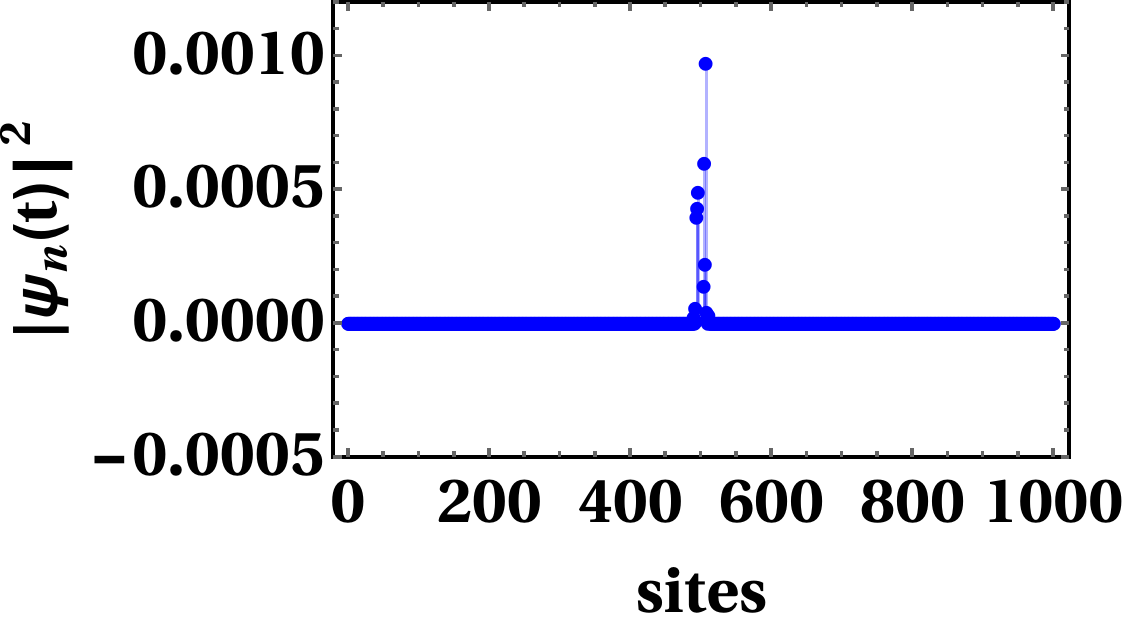}
(g)\includegraphics[width=.44\columnwidth]{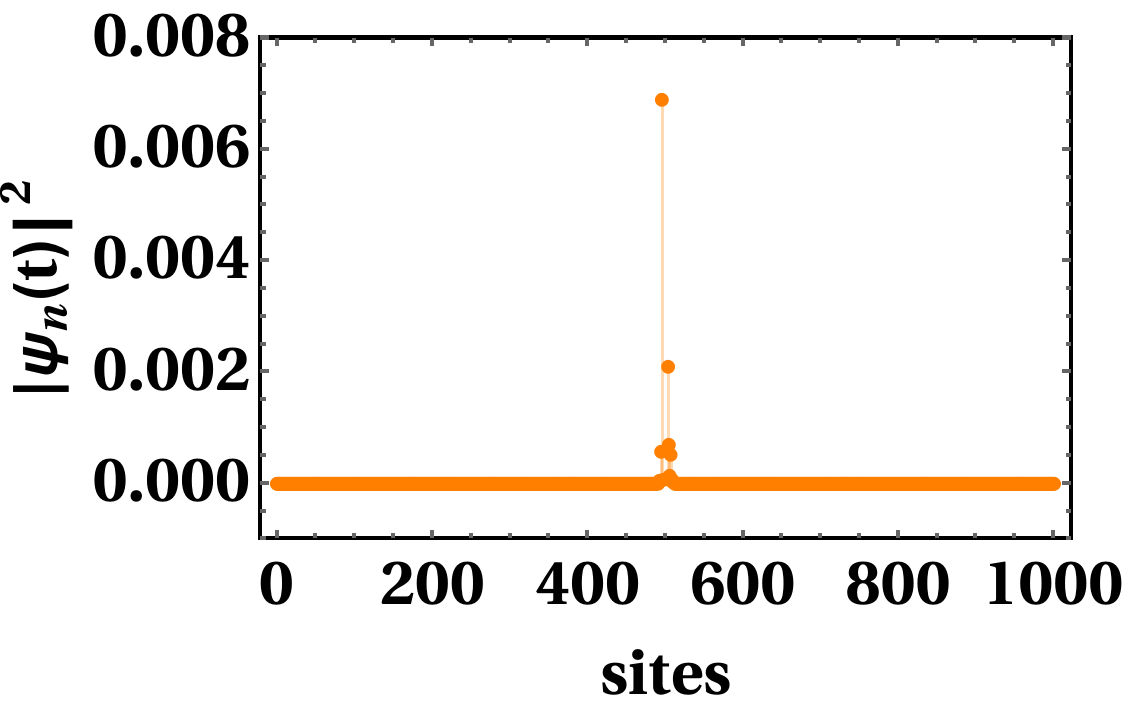}
(h)\includegraphics[width=.44\columnwidth]{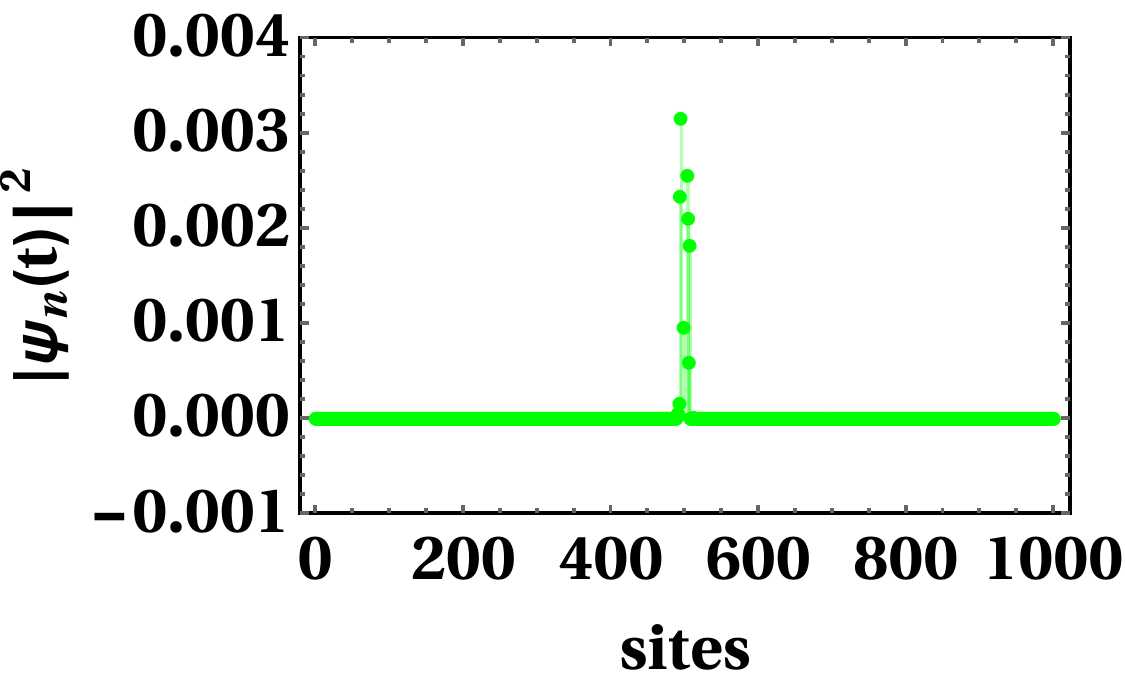}
(i)\includegraphics[width=.44\columnwidth]{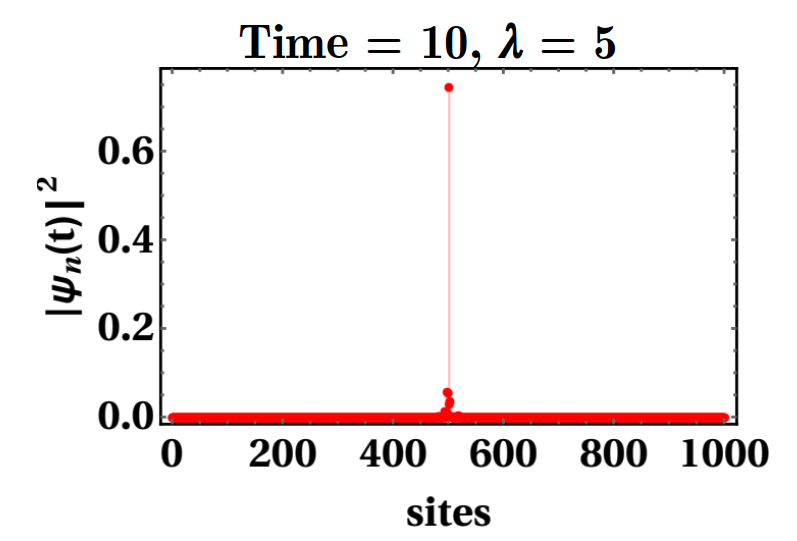}
(j)\includegraphics[width=.44\columnwidth]{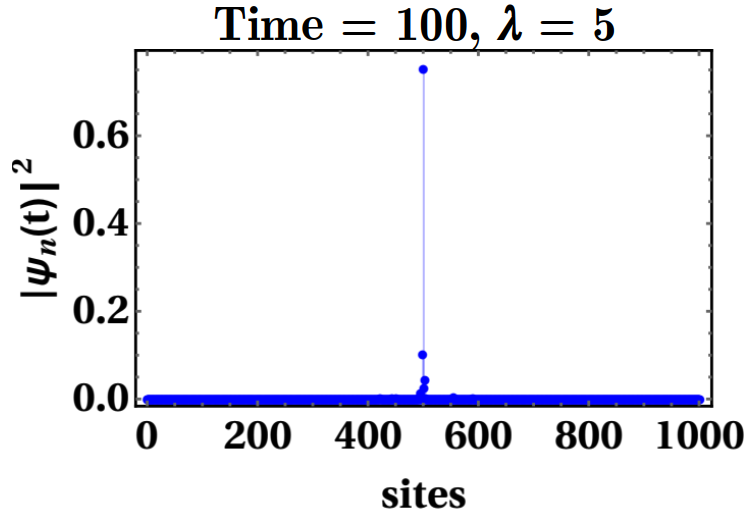}
(k)\includegraphics[width=.44\columnwidth]{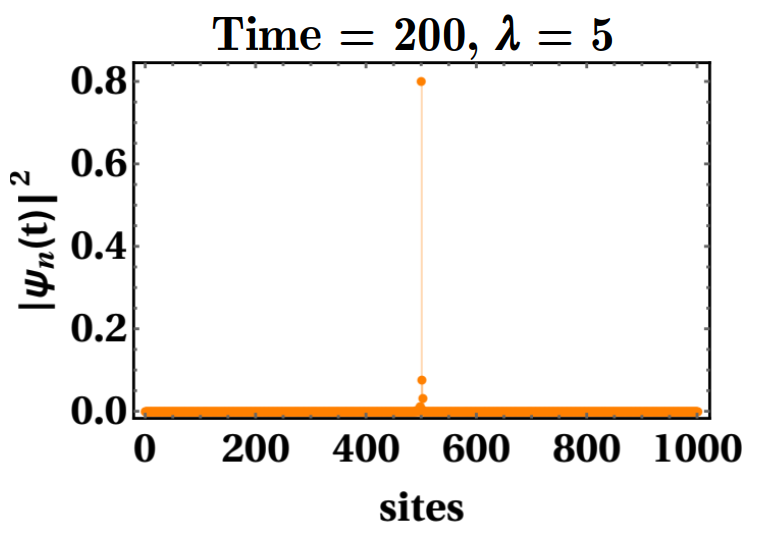}
(l)\includegraphics[width=.44\columnwidth]{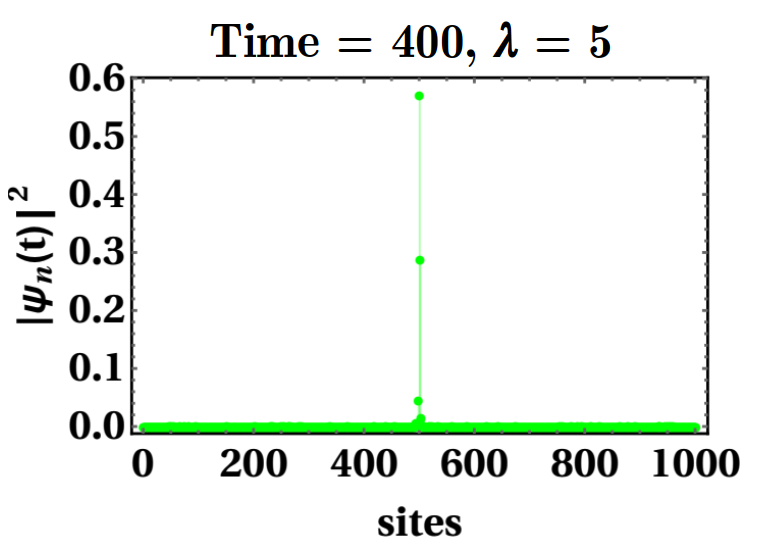}
(m)\includegraphics[width=.44\columnwidth]{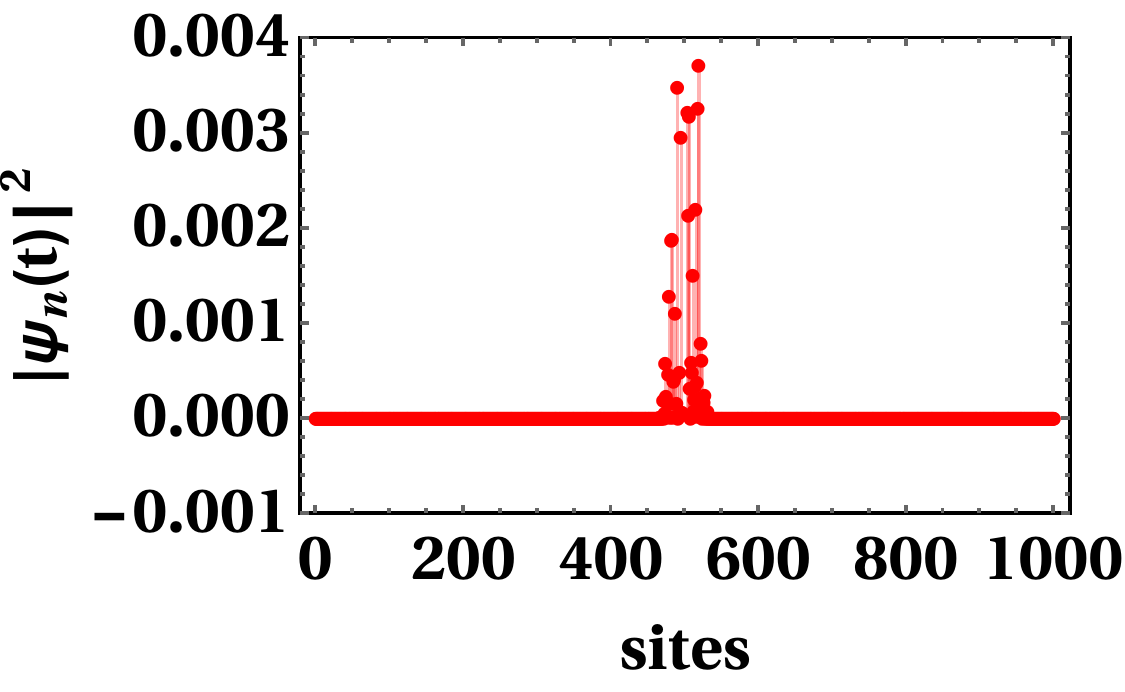}
(n)\includegraphics[width=.44\columnwidth]{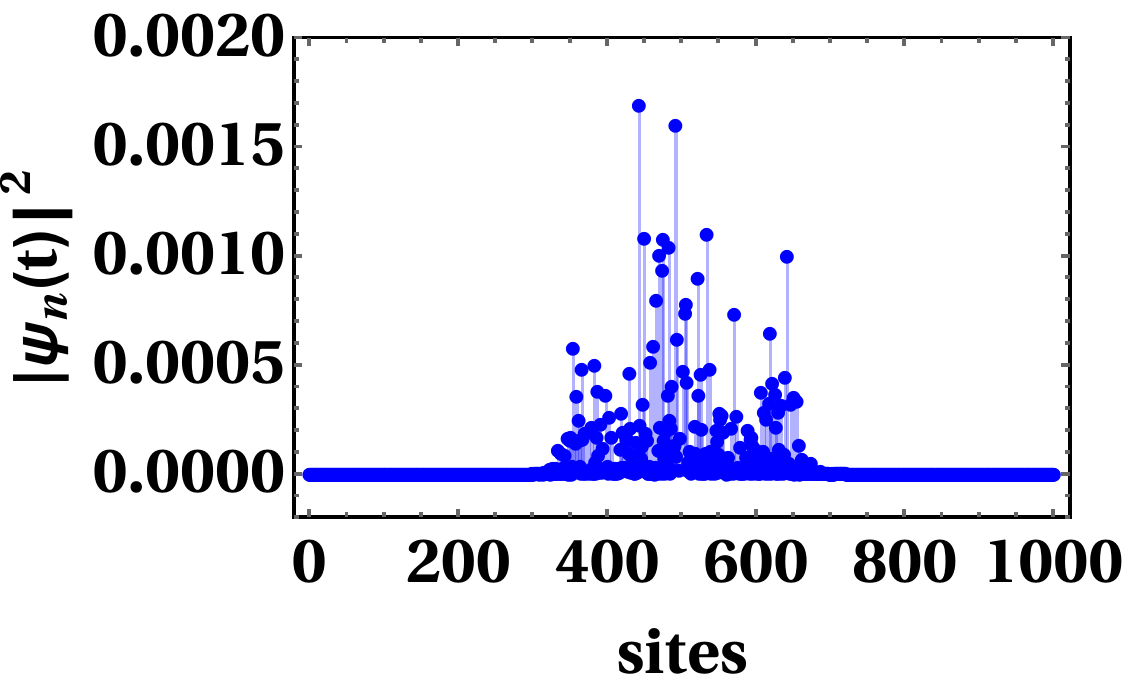}
(o)\includegraphics[width=.44\columnwidth]{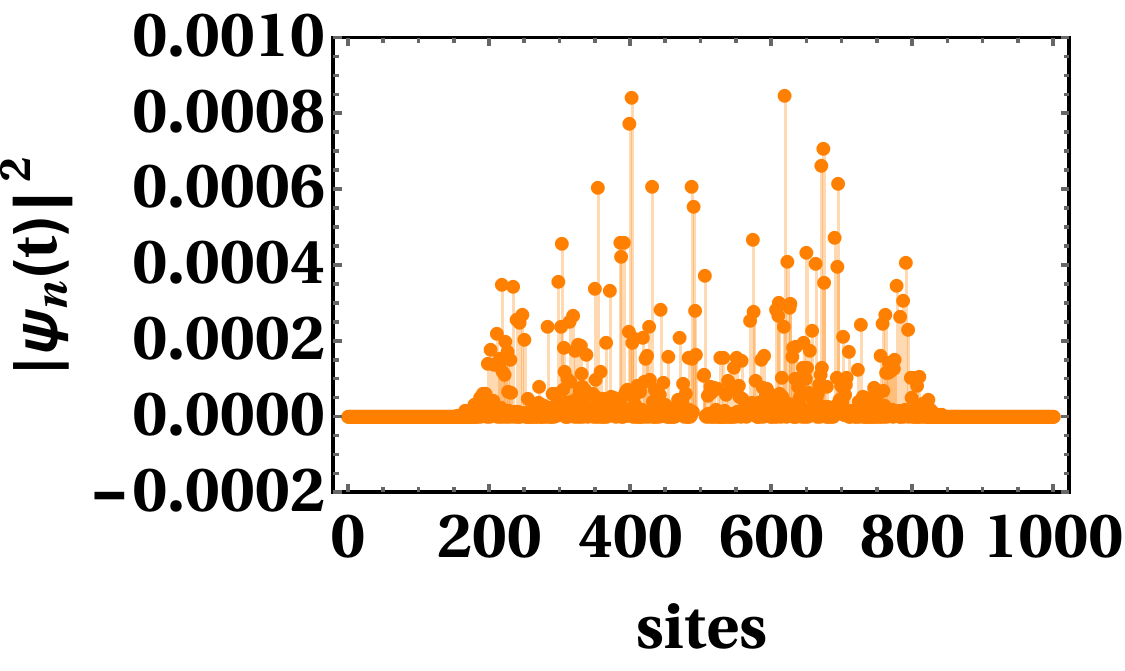}
(p)\includegraphics[width=.44\columnwidth]{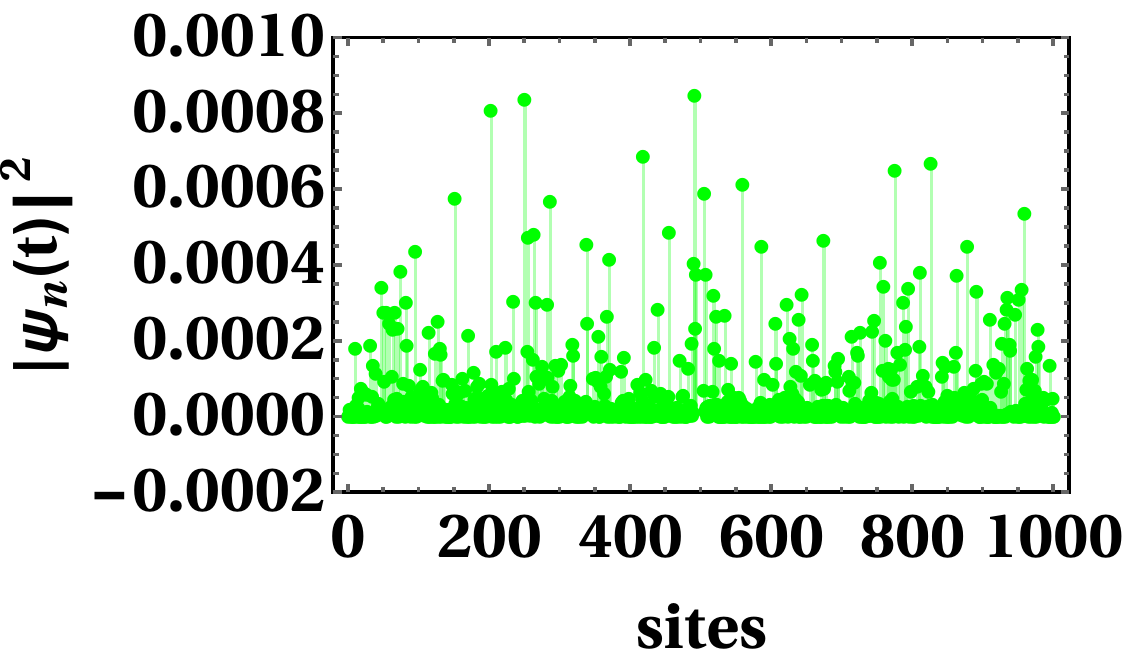}
\caption{(Color online) The variation of the spatial distribution of the wave packet for (a-h) ladder network ($1002 sites$), and (i-p) ultrathin graphene nano-ribbon network ($1002 sites$)  with $time = 10$(red colored), $time = 100$(blue colored), and $time = 200$(orange colored) and $time = 400$(green colored) respectively. Initially, we put the wave packet on the $501^{th}$ site. The rung hopping parameter is given by $\Gamma_{n}= y (1+ \lambda \cos~( 2 \pi Q n))$, where $Q = \frac{\sqrt{5}-1}{2}$. The strength of the modulation is chosen as $\lambda = 5$. The onsite potential $\epsilon$ is chosen as zero. The parameter $ y $ is set as $1$. (e,f,g,h) and (m,n,o,p) are the magnified versions of (a,b,c,d) and (i,j,k,l) respectively.}
\label{psin3}
\end{figure*}

\subsection{Multifractal Analysis}
Now we will discuss the finite size scaling of MIPR for different system sizes $N$ when the systems are in their critical region ( $\lambda = 2$ for the ladder model and $\lambda = 0.75$ for the ultrathin GNR network). The corresponding graphical presentation for these quantum networks is shown in Fig.~\ref{scaling}. For the delocalized region, MIPR decays with the system size as $N^{-1}$, whereas MIPR has no dependency on the system size ($N^0$) when the system is in its localized region~\cite{Joana}. A special behaviour of MIPR with system size is observed in its critical region (near transition point $\lambda_c$). The decay exponent of MIPR in this particular case turns out to be any value between $0$ and $-1$. In our case, it decays as $N^{-0.50}$ (Red colored) for the ladder network and $N^{-0.69}$ (blue colored) for the ultrathin GNR network. That's why this region is termed as a multifractal region~\cite{Joana}.\\

One way to represent the multifractal behavior is with the generalized inverse participation ratio ($\mathcal{I}_q$), which is defined as~\cite{Evers,Hiramoto,Amrita1,Sajid},
\begin{equation}
    \mathcal{I}_q = \frac{\sum_n |\psi_n|^{2q}}{(\sum_n |\psi_n|^2)^q}
\end{equation}
Here $q$ is the moment (an integer). This $\mathcal{I}_q$ exhibits anomalous scaling behavior with the system size $N$, via $\mathcal{I}_q \approx N^{-\tau_q}$, where $\tau_q = D_{q}(q-1)$. This $D_q$ is the fractal dimension~\cite{Evers}. For a pure delocalized and localized state $D_q$ is $0$ and $1$ respectively irrespective of any value of moment $q$. Apart from this, when $D_q$ turns out as a particular constant value between $0$ and $1$ for all values of $q$, we can say that the state is fractal with one fractal dimension. On the other hand, the $D_q$ is in between $0$ and $1$, but depend on the moment $q$, such states are called multifractal.\par
Now, we investigate the multifractal behaviour of the eigenstates in the critical region (at the transition point or very close to it) $\lambda = 2$ and $0.75$ for these quantum networks respectively. Fig.~\ref{multi}(a,c) represents the distribution of multifractal wavefunction at the ground state energy (a) $E = -3.59751$ (ladder network), and (c) $E = -2.7061$ (ultrathin GNR network)  respectively. Corresponding fractal dimension $D_q$ is plotted against $q$ in Fig.~\ref{multi}(b, d), and dependency with $q$ is apparent for the quantum networks. Both cases $D_{q}$ decay with the moment $q$. So from the special amplitude distribution of the wavefunction and moment ($q$) dependent fractional values of fractal dimension ($D_{q}$), it is apparent that both networks exhibit a multifractal nature in their critical region.

\section{Quantum Dynamics}

At time $t = 0$, the electron's wave packet is expressed as,
\begin{equation}
\ket{\Psi(0)} = \sum_{n} C_{n} (0) \ket{\psi_{n}(0)} 
\end{equation}
The initial probability amplitude for finding a particle at the $n^{th}$ site is denoted by $C_n$. When the particle is initially released at the $ m^{th}$ site, this amplitude is given by $C_n (0) = \bra{\psi_{n}(0)}\ket{m}$. After some time, the state evolves as,
\begin{equation}
  \ket{\Psi(t)} = \sum_{n} C_n e^{-i E_{n} t} \ket{\psi_{n}(0)}
 \label{psit}
\end{equation}
We now study the fundamental quantities that govern the quantum dynamics of a wave packet on finite-sized systems with the geometries shown in Fig.~\ref{figure}.


\subsection{Spreading of the wave packet}
We measure the spatial variation of the wave packet $ |\psi_{n}(t)|^2 $ at different instants of time. At first, we calculated the time-evolved states for ladder and ultrathin GNR networks using Eq.~\ref{psit}. We choose $1002$ atomic sites and initially put the wave packet at $ m = 501^{th}$ site (almost in the middle of the systems) throughout all calculations related to the quantum dynamics.\par  Fig.~\ref{psin1}(a,c,e) and (b,d,f) are for the ladder and the ultrathin GNR network, which shows the spatial distribution of the wavepacket with different time, $t = 10 $, $ t = 100$ and $ t = 200$ respectively when both systems are in its delocalized region $\lambda = 0.25$ ($\lambda < \lambda_{c})$. All eigenstates corresponding to this region are fully delocalized in nature, with the increment of time the wave packet spreads out from its initial location $m$. It is observed from all graphs (Fig.~\ref{psin1}), that amplitude of $ |\psi_{n}(t)|^2 $ becomes lower (still non-zero) with time evolution. This is because the total probability is always $1$. We will discuss further that, these lower values of the amplitudes of $ |\psi_{n}(t)|^2 $ can take an important role in mean square displacement (MSD) calculation.\par

Fig.~\ref{psin2} demonstrates the spatial distribution of wavepacket with time when the systems are in the critical region ($\lambda = 2$ for ladder network and $\lambda = 0.75 $ for ultrathin GNR network). Fig.~\ref{psin2}(a-d) is for the ladder model with different time instants which indicates the rate of spreading of the wavepacket is small (please see the corresponding magnified versions Fig.~\ref{psin2}(e-h)) comparatively from its delocalized region. The spatial variation of the wavepacket within the critical region ($\lambda =0.75$) for the ultrathin GNR network is shown in  Fig.~\ref{psin2}(i-l). In this case, the wavepacket spreads out all over the system after a finite time (please see the magnified version Fig.~\ref{psin2}(m-p)).\par

The distribution of the wavepacket of the ladder network in its localized region ($\lambda = 5$) is shown in Fig.~\ref{psin3}(a-d) (corresponding magnified versions are depicted in Fig.~\ref{psin3}(e-h)). Due to all states being localized in this region, the wavepacket is always in its initial site of release, no matter how much the system evolves with time. Fig.~\ref{psin3}(i-l) is for the ultrathin GNR network with $\lambda = 5$. In this situation, both states are present in the spectrum viz, the central energy region is delocalized and two sides of the central energy region get localized. Here $\psi_{n}(t)$ contains both eigenstates (localized and delocalized). Now with the increment of time, the delocalized part of the wavepacket spreads out (please see Fig.~\ref{psin3}(m-p)) and the localized part of the wavepacket remains in its initial location (please see Fig.~\ref{psin3}(i-l)). Although the probability amplitudes of the delocalized part of the wavepacket are very small, still play an important role in measuring the mean square displacement (MSD).

\subsection{Mean Square Displacement}

The spreading of the wave packet with time, which was initially at a particular site $m$ (say) at $t = 0$ is measured by the mean square displacement (MSD). It is defined as~\cite{katsanos},
\begin{equation}
    \sigma^{2}(t) = \sum_{n} (n - m)^2 |\psi_{n}(t)|^2
    \label{msdeq}
\end{equation}
The behaviour of $\sigma^{2}(t)$ shows an asymptotic dependence on time in the long time limit. This time dependency is described by a power law~\cite{katsanos} form $\sigma^{2}(t) \sim t^{\mu}$. Depending upon the numerical values of this exponent $\mu$, the dynamics of wave packets can be classified into different regions, like localized for $\mu=0$, and sub-diffusive for $0 < \mu <1$~\cite{siegle,vollmer,bodrova}. Whereas $\mu = 1$, $2 > \mu > 1 $, and $\mu = 2$ indicate ordinary diffusion, super diffusion, and ballistic motion respectively~\cite{katsanos,arka,sougata}.\\

\begin{figure}[ht]
\centering
(a)\includegraphics[width=.8\columnwidth]{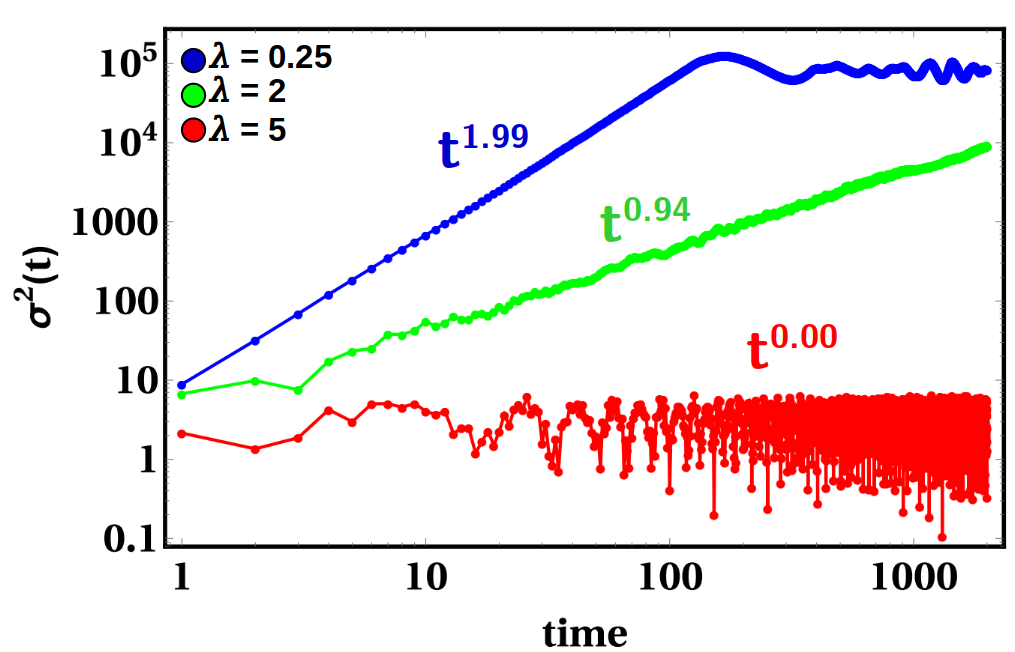}
(b)\includegraphics[width=.8\columnwidth]{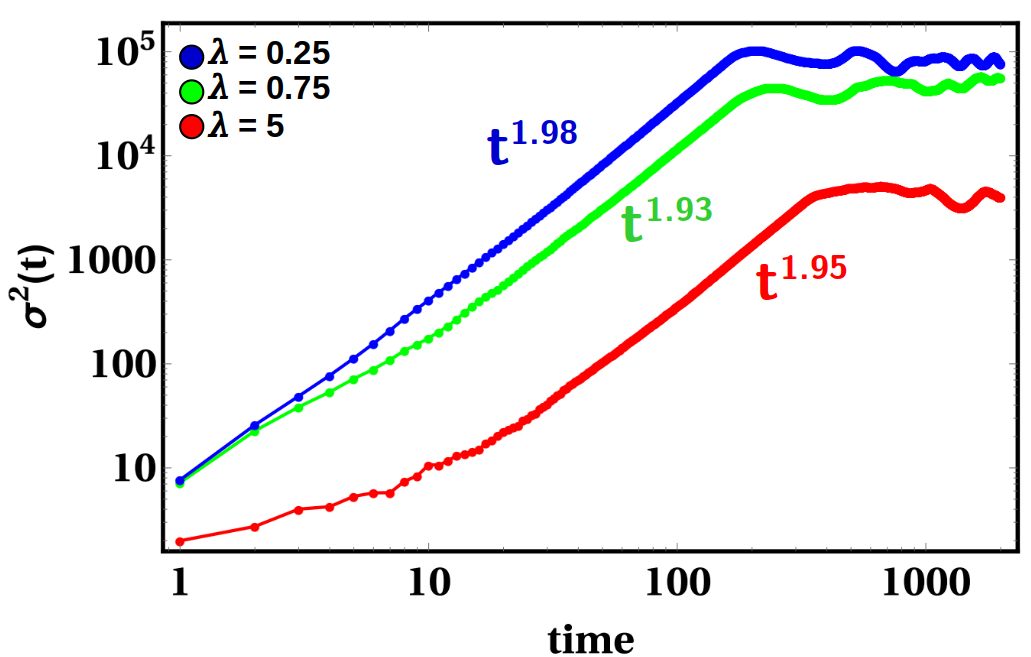}

\caption{(Color online) The mean-square displacement of a wave packet as a function of time (in a log-log scale) for (a) ladder network ($1002$ sites), (b) ultrathin graphene nano-ribbon network ($1002$ sites). Initially, we put the wave packet on the $501^{th}$ site. The rung hopping parameter is given by $\Gamma_{n}= y (1+ \lambda \cos~( 2 \pi Q n))$, where $Q = \frac{\sqrt{5}-1}{2}$. The onsite potential $\epsilon$ is chosen as zero. The parameter $ y $ is set as $1$. }
\label{msd}
\end{figure}


\begin{figure}[ht]
\centering
\includegraphics[width=.8\columnwidth]{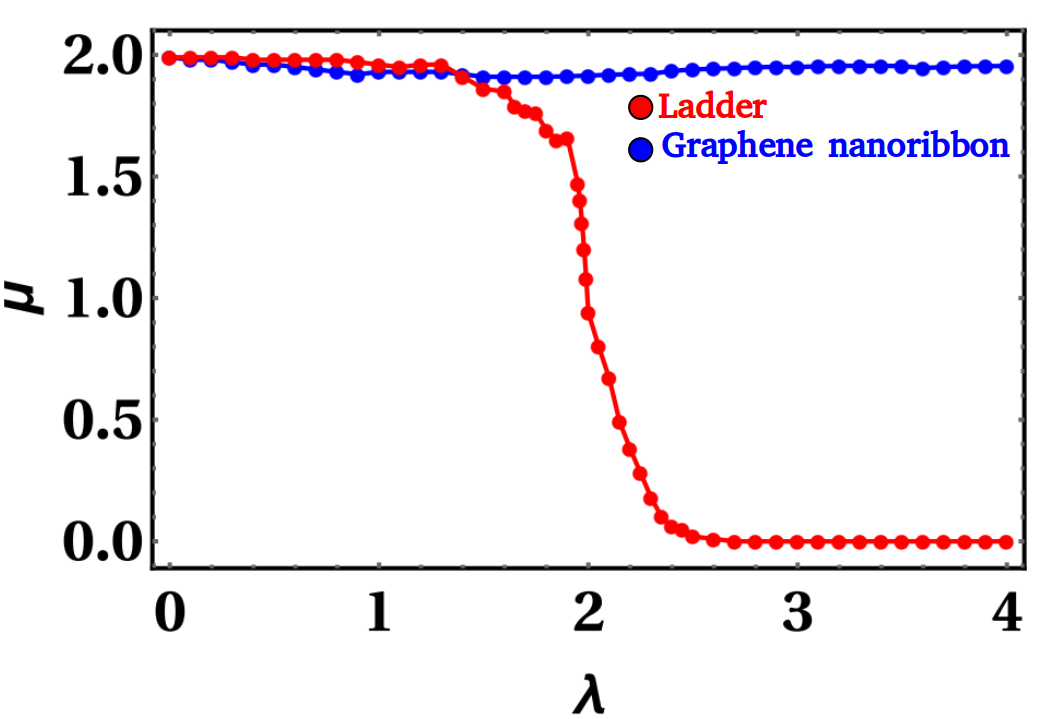}

\caption{(Color online)  Variation of the exponent $\mu$ of the mean square displacement with the strength of the AAH modulation $\lambda$ for ladder network (red colored) and ultrathin graphene nano-ribbon network (blue colored) containing $1002$ atomic sites. Initially, we put the wave packet on the $501^{th}$ site. The rung hopping parameter is given by $\Gamma_{n}= y (1+ \lambda \cos~( 2 \pi Q n))$, where $Q = \frac{\sqrt{5}-1}{2}$. The onsite potential $\epsilon$ is chosen as zero. The parameter $ y $ is set as $1$. }
\label{time-exponent}
\end{figure}

We have calculated means square displacement (MSD) for the ladder network (containing $1002$ atomic sites) for different values of the Aubry modulation strength $\lambda$ using Eq.~\ref{msdeq} which is depicted in Fig.~\ref{msd}(a). Initially (at $t =0$) we put the wave packet $m = 501 ^{th}$ site. Now by calculating MSD, we can measure the spreading of the wave packet from its initial location with time. When the strength of AAH modulation $\lambda = 0.25$, all eigenstates become extended in nature. The MSD varies with time as $t^{1.99}$ (blue colored). The system exhibits its ballistic nature, which is expected.  Further increment of $\lambda$, say $\lambda = 2$ the system enters its critical region and the corresponding long time behaviour of MSD varies as $t^{0.94}$ (green colored). So long-time behavior of the wave packet in its critical region is sub-diffusive or very close to the ordinary diffusive motions. At $\lambda = 5$ the spreading of the wave packet is very small and the time exponent ($\mu$) becomes $0$ (red colored), such that the system exhibits localized character.  So from the long-time behaviour of the wave packet of the ladder model, it is clear that the system shows ballistic-super(ordinary, sub)diffusive-localization character, depending upon the strength of the modulation $\lambda$. This result again supports the occurrence of metal-to-insulator phase transition in the ladder network under AAH modulation.

Now we discuss the interesting behavior of quantum dynamics of the ultrathin graphene nano-ribbon networks. Fig.~\ref{msd} (b) shows the variation of MSD for the ultrathin GNR network with different modulation strengths $\lambda$. The graphs show that the long-time behavior of MSD with time is always very close to the ballistic motions, no matter how much strength $\lambda$ of the modulation is applied. To investigate the proper reason behind this, back to Eq.~\ref{psit}. The co-efficient $C_n$ here plays an important role in controlling the Dynamics of the finite MIPR region as shown in Fig.~\ref{mipr}(b). For a completely delocalized region, all wavefunctions corresponding to any energy eigenvalue are distributed all over the system. As a result, most of the  $C_n$ are non-zero. Similarly when a particular region is localized corresponding wave functions live in only a small number of sites and in all other positions have no amplitude. Such that a small number of $C_n$ are non-zero and all others are zero. On the other hand, for a mixed region (both localized and delocalized states), the maximum $C_n$ that comes out from the localized eigenstates becomes zero, and almost all $C_n$ corresponding to the delocalized states remain finite. The non-zero $C_n$ that comes from both localized and delocalized states is entered in the calculation of $\psi(t)$ (see Eq.~\ref{psit}). Due to this $\psi(t)$ contains both localized and de-localized parts.  At time $t = 0$, the $\psi_{n}(t)$ is only its initial site of release (here $m$). When time is increased the delocalized part of the wavepacket spreads out, whereas the localized part remains in its initial position $m$. But this amplitude of the wavepacket which remains in its initial location ($m$) never enters the MSD calculation because $(n-m)^2 = 0$ (see Eq.~\ref{msdeq}) at $n = m$. As discussed in the previous part, although $ |\psi_{n}(t)|^2 $ contains a very small value (see Fig.~\ref{psin3} (m, n, o, p)) still have a finite MSD because the term $(n-m)^2$ turns out as a quite large value when it moves away from its initial site of release. So for a mixed state, MSD is mainly controlled by the delocalized part of the wavepacket. As the central energy region of the ultrathin GNR network is always delocalized irrespective of $\lambda$, the MSD still varies as $~t^2$ (or very close to it) at higher modulation strength.\par
Fig.~\ref{time-exponent} demonstrates the behaviour of the exponent of MSD for both these quantum networks. The exponent is $2$ when no modulation is applied $\lambda = 0$, which is obvious. When the applied strength of the modulation $\lambda$ is increased, the exponent goes from $2$ to $0$ (by crossing super-ordinary-sub diffusive region) for the ladder network (red colored). So the metal-to-insulator phase transition is confirmed again. Interestingly, for the ultrathin GNR network, the exponent always turns out very close to $2$ with any value of applied strength (blue colored). This is due to the extendedness of the central energy region (already discussed earlier).\par
\subsection{Return Probability}
If a wave packet is initially released at position $m$ at time $t = 0$, the probability of finding the wavepacket at the same position $m$ at a later time $t$ is denoted by $P_{m}(t) = |\psi_{m}(t)|^2$. This is referred to as the return probability (RP). If, after a long time duration, the RP at the initial position 
$m$ remains nonzero, it indicates localization. On the other hand, in a delocalized phase, the RP approaches zero as time progresses~\cite{guan}.

 \begin{figure}[ht!]
\centering
(a)\includegraphics[width=.44\columnwidth]{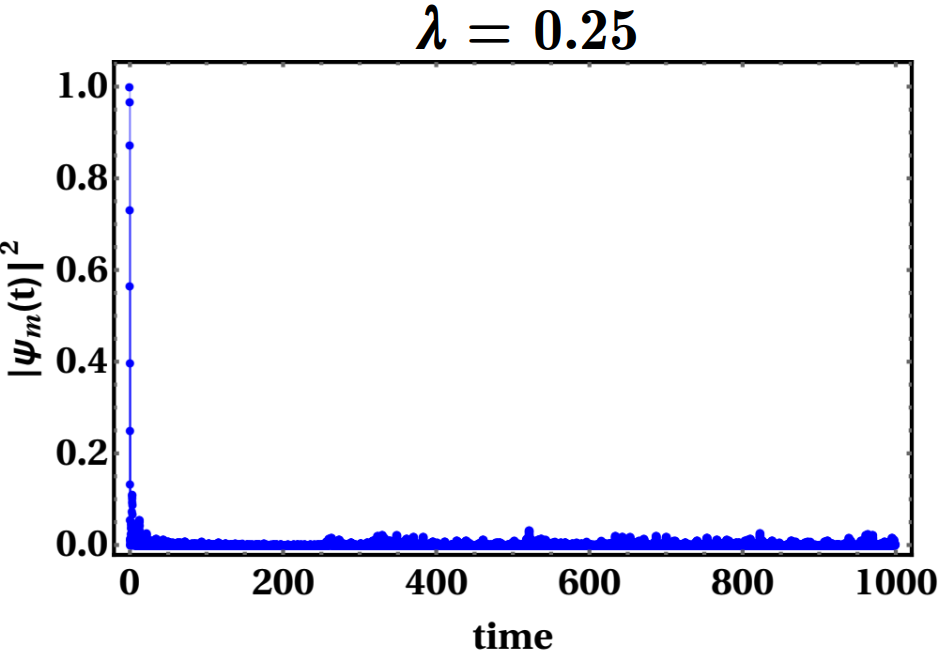}
(b)\includegraphics[width=.44\columnwidth]{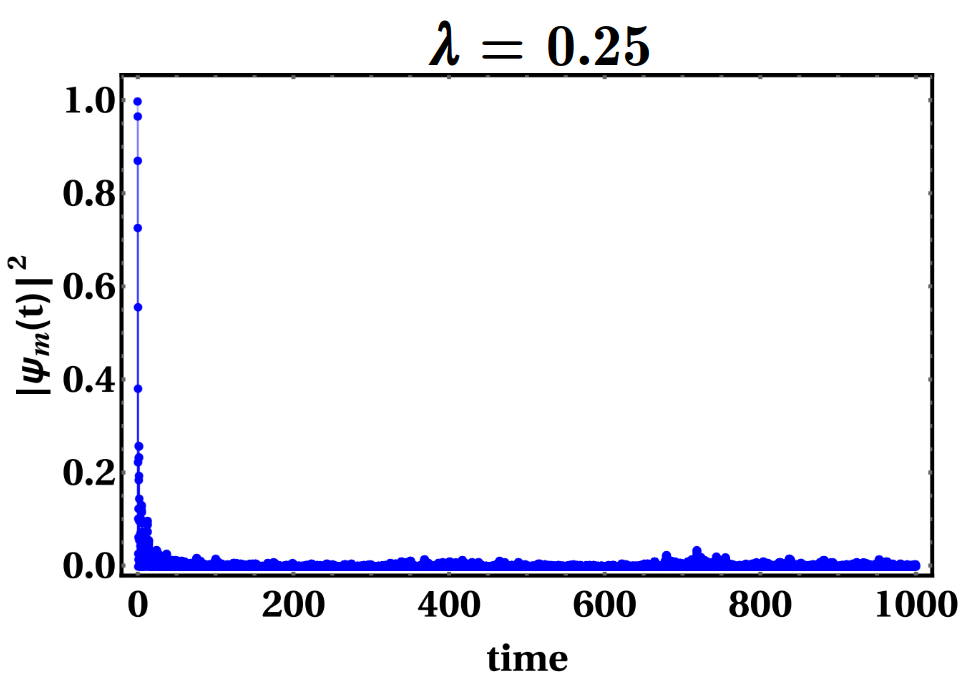}
(c)\includegraphics[width=.44\columnwidth]{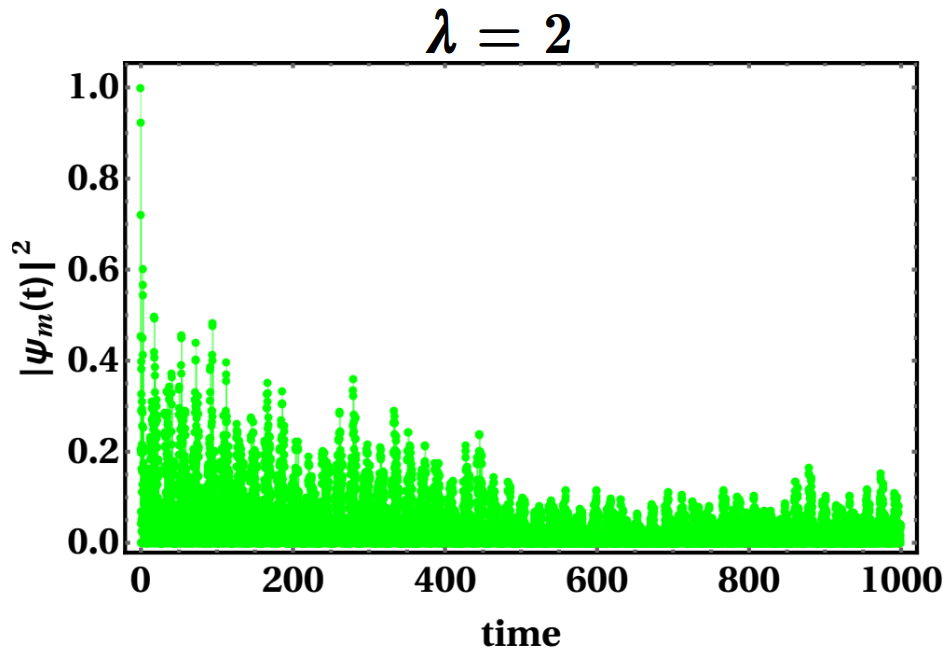}
(d)\includegraphics[width=.44\columnwidth]{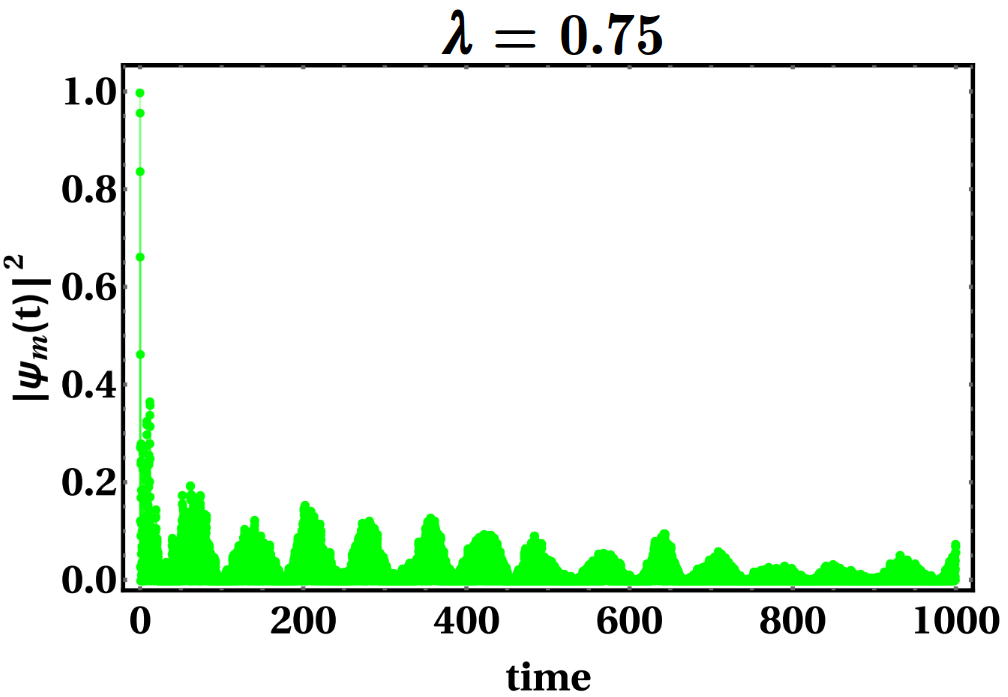}
(e)\includegraphics[width=.44\columnwidth]{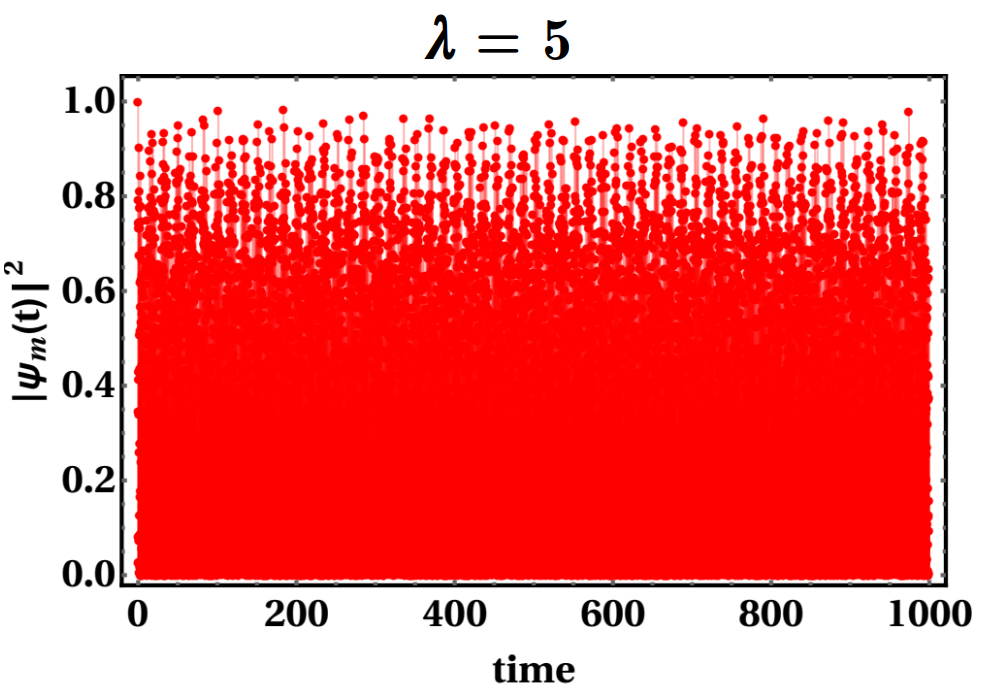}
(f)\includegraphics[width=.44\columnwidth]{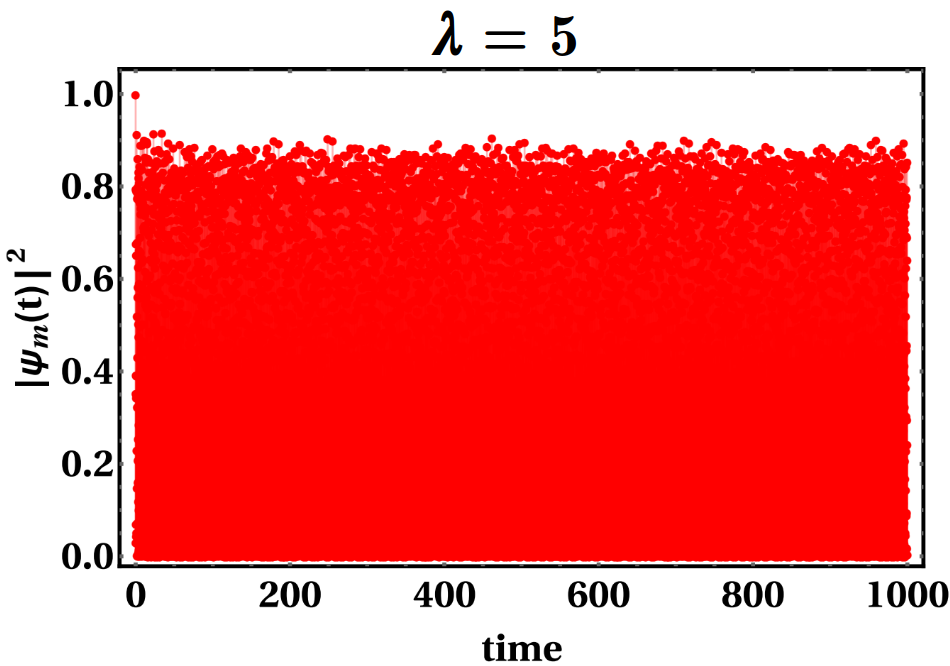}
\caption{(Color online) The probability of finding the wavepacket at the initial site as a function of time for (a, c, e) ladder network ($1002$ sites) and (b, d, f) ultrathin graphene nano-ribbon network ($1002$ sites) with different values of $\lambda$. Initially, we put the wave packet on the $501^{th}$ site. The rung hopping parameter is given by $\Gamma_{n}= y (1+ \lambda \cos~( 2 \pi Q n))$, where $Q = \frac{\sqrt{5}-1}{2}$. The onsite potential $\epsilon$ is chosen as zero. The parameter $ y $ is set as $1$.}
\label{RP}
\end{figure}

Fig.~\ref{RP} (a, c, e) shows the return probability of the ladder model with various modulation strengths $\lambda$. We calculated the return probability in its three different regions, $\lambda = 0.25$ (blue-colored, de-localized region), $\lambda = 2$ (green-colored, critical region), and $\lambda =5$ (red-colored, localized region). For delocalized region $\lambda = 0.25$ the wavepacket spreads out from its initial location and the return probability goes to zero with time (see Fig.~\ref{RP} (a)). At $\lambda = 2$, the system enters into its critical region and shows ordinary-diffusive motion (as discussed in the MSD section). The corresponding return probability decays with time and goes to zero after a large time evolution (see Fig.~\ref{RP} (b)). Fig.~\ref{RP} (c) shows the behavior of return probability with time for its localized region, as the wave packet is always in its initial site of release, the return probability remains finite no matter how much the system evolves with time.\par
The return probability for the ultrathin graphene nano-ribbon network in its three different regions, $\lambda = 0.25$ (blue-colored), $\lambda = 0.75$ (green-colored), and $\lambda =5$ (red-colored) are depicted in Fig.~\ref{RP} (b, d, f). Now for $\lambda = 0.25$, all states are de-localized and the return probability becomes zero as the time flows on (see Fig.~\ref{RP} (b)).  Fig.~\ref{RP} (d) shows the return probability in its critical region which decays to zero after a long time. These are obvious.  As already discussed in the MSD section, for the mixed states the MSD always varies close to $t^2$ because the amplitude of the wavepacket at its initial site of release is never entered in the MSD calculation. But in RP measurement, we mainly calculate the probability of finding the wavepacket at the initial site $m$. With time increment, for the mixed states, the delocalized part of the wavepacket spreads out but the localized part of the wavepacket remains its initial location after a long time. As a result system shows a finite return probability.  Fig.~\ref{RP} (f) shows the behavior of return probability with time for its mixed region, as the localized part of the wave packet is always in its initial site of release, the return probability remains finite.\par
So for the ultrathin GNR network, although the MSD behaviour is always ballistics in nature, at the same time it has a finite return probability with higher modulation strength $\lambda$.\par
Now, we measure another popular measurement of quantum dynamics, the temporal autocorrelation function (TAF). This is a time-averaged representation of the return probability, which can be defined as~\cite{katsanos,arka}.
\begin{equation}
    C(t) = \frac{1}{t}  \int_{0}^{t} |\psi_{m}(t)|^2 \,dt 
    \label{eq-taf}
\end{equation}\\
The behaviour of TAF for large times is described by the power-law~\cite{katsanos} $C(t) = t^{\delta}$. For the pure de-localized case, this exponent $\delta$ is $\approx - 1$ in the long time limit. Whereas it almost goes as $t^{0}$ for the localized system.

\begin{figure}[ht!]
\centering
(a)\includegraphics[width=.44\columnwidth]{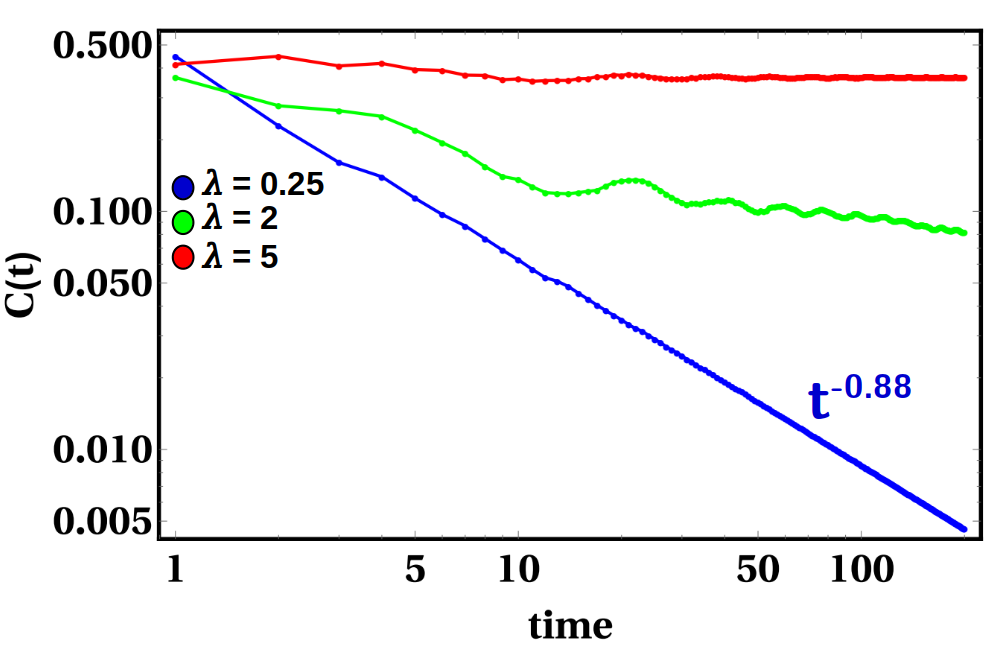}
(b)\includegraphics[width=.44\columnwidth]{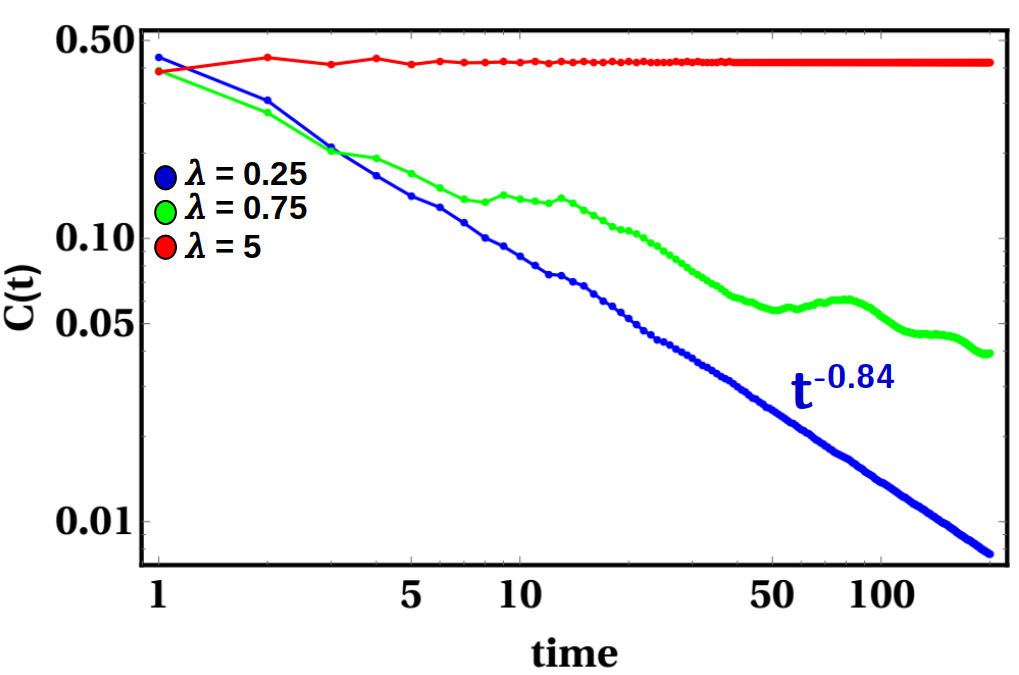}

\caption{(Color online) The temporal auto-correlation function is plotted against time using a log-log scale, for the (a) ladder network ($1002$ sites) and (b) ultrathin graphene nano-ribbon network ($1002$ sites) with different modulation strength $\lambda$. Initially, we put the wavepacket on the $501^{th}$ site. The rung hopping parameter is given by $\Gamma_{n}= y (1+ \lambda \cos( 2 \pi Q n))$, where $Q = \frac{\sqrt{5}-1}{2}$. The onsite potential $\epsilon$ is chosen as zero. The parameter $ y $ is set as $1$. For (a) $\lambda$ is chosen as $\lambda = 0.25$ (blue colored), $\lambda = 2$ (green colored), $\lambda = 5$ (red colored) and for (b) $\lambda = 0.25$ (blue colored), $\lambda = 0.75$ (green colored), $\lambda = 5$ (red colored). }
\label{taf}
\end{figure}
In Fig.~\ref{taf}(a) TAF is plotted for the ladder network. For $\lambda = 0.25$(de-localized region) TAF decays out with time (blue colored) and the corresponding exponent is $\delta \approx -0.88$, going to $\approx -1$ after a long time evolution. The TAF in the critical region $\lambda = 2$ shows a small decay rate as it is sub(or close to ordinary)-diffusive in nature (green colored). At $\lambda = 5$, all eigenstates get localized and the wavepacket remains in the initial site of release. So no decay is observed and it goes as $t^{0}$ (red colored).\par
Variation of TAF of the ultrathin GNR network is depicted in Fig.~\ref{taf}(b). For the de-localized region $\lambda = 0.25$, it decays out and the exponent becomes $\approx - 0.84$ (blue colored). Similarly, The decay of TAF with time is small for the critical region $\lambda = 0.75$ (green colored). When $\lambda = 5$, both localized and delocalized states are present in the spectrum. As discussed earlier the localized part of the wavepacket never spread out with time, the TAF basically has no decay and exponent $\delta \approx 0$ (red colored).\par
\subsection{Time evolution of Inverse Participation Ratio}
In this section, we investigate the nature of the inverse participation ratio (IPR) when the system evolves with time. The inverse participation ratio is defined as~\cite{Mauro,sougata},
\begin{equation}
    IPR(t) = \sum_{n} |\psi_{n}(t)|^4
\end{equation}
Here the atomic site index is denoted by $n$. If we release the wave packet at $t=0$ on its initial site $m$, then only $ |\psi_{m}(t)|^4  = 1$, and in all other positions, it goes zero. As a result, $IPR(0)$ is always unity for all states (delocalized, localized, or mixed states).

\begin{figure}[ht!]
\centering
(a)\includegraphics[width=.44\columnwidth]{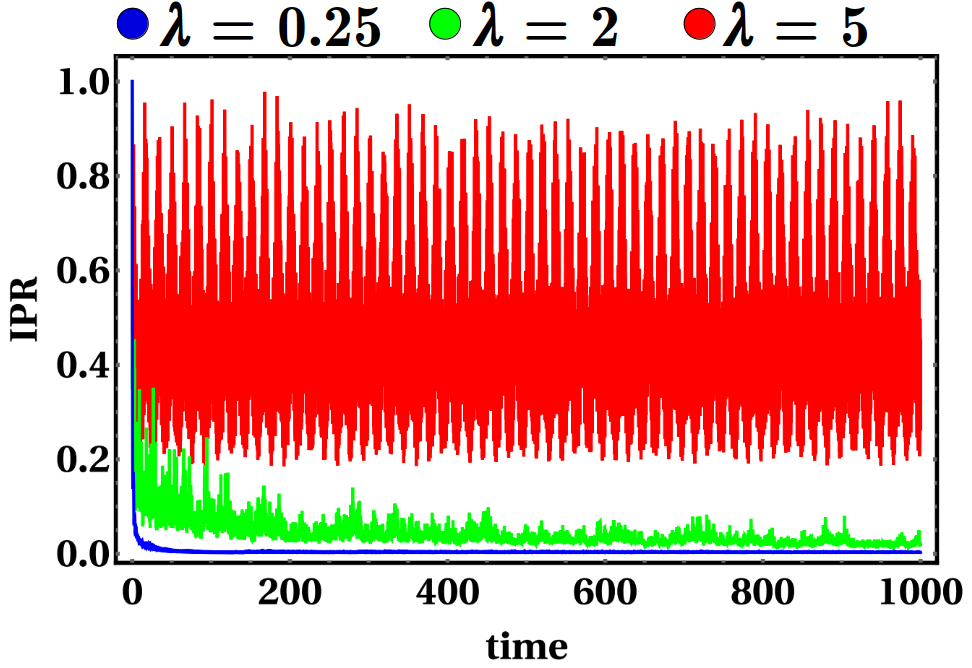}
(b)\includegraphics[width=.44\columnwidth]{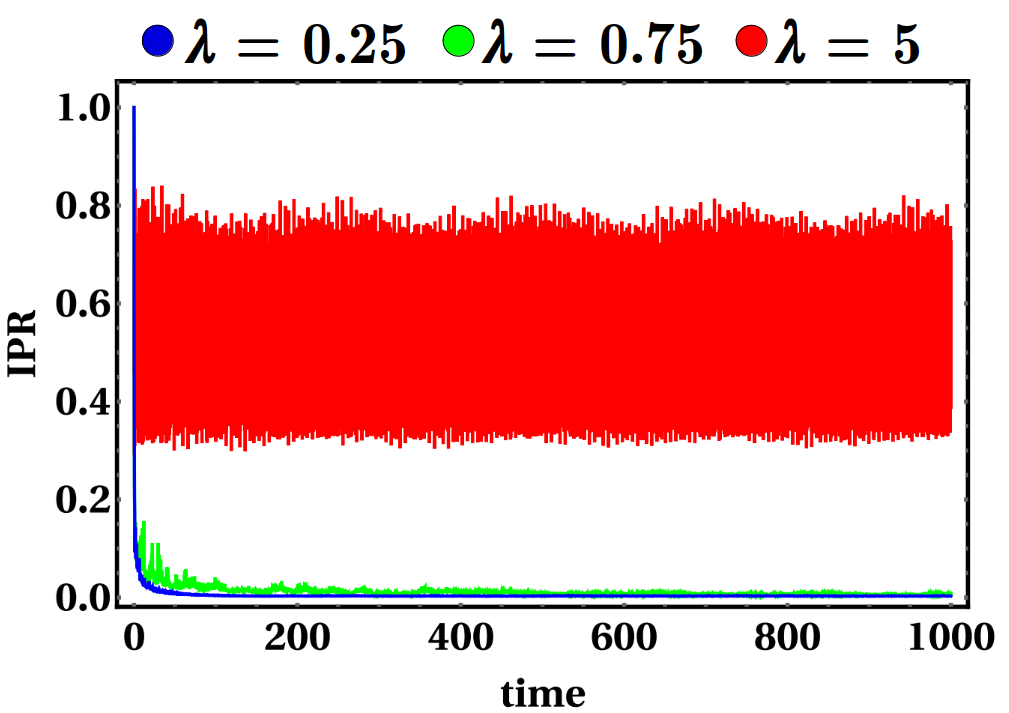}

\caption{(Color online) Variation of IPR as a function of time for (a) ladder network ($1002$ sites) and (b) ultrathin graphene nano-ribbon ($1002$ sites) with different values of $\lambda$. Initially, we put the wave packet on the $501^{th}$ site. The rung hopping parameter is given by $\Gamma_{n}=y (1+ \lambda \cos~( 2 \pi Q n))$, where $Q = \frac{\sqrt{5}-1}{2}$. The onsite potential $\epsilon$ is chosen as zero. The parameter $ y $ is set as $1$. }
\label{iprt}
\end{figure}

The wavepacket spreads out from its initial location with a non-zero probability $|\psi_{n}(t)|^2$ on each site while delocalizing with time. Over time, the amplitude of this probability decreases. Thus, when time grows sufficiently large, the sum of $|\psi_{n}(t)|^4$ over all atomic sites tends to zero.
Nevertheless, the wavepacket is limited to the area surrounding its starting position during a localized phase. The probability $|\psi_{n}(t)|^2$ is almost zero at all other locations, but it is larger near its starting location. $IPR(t)$ never decays to zero because of this kind of distribution, regardless of how long the dynamics are observed.
Fig.~\ref{iprt}(a) shows the variation of IPR with time for a ladder network when different modulation strengths $\lambda$ are applied. For $\lambda = 0.25$, IPR goes to zero when the system evolves with time (Blue colored). At critical region $\lambda = 2$, the system shows sub-diffusive motion and the corresponding IPR contains small finite values and goes to zero after a long time(green colored). Whereas with time evolution the IPR  gives a non-zero value (with an oscillatory character) when the system enters its localized region $\lambda = 5$ (Red colored) and never decays with time.\par
The variation of IPR with time for the ultrathin GNR network is shown in Fig.~\ref{iprt}(b). It becomes zero just when time is switched on at $\lambda = 0.25$ (blue-colored) as all eigenstates are delocalized. The IPR in its critical region decays to zero after a finite time (green colored). For the mixed region (say $\lambda =5$), the localized part of the wavepacket is always strict in its initial position and the delocalized part spreads out with time evolution. As a result, IPR always contains finite values (red colored) still after a large time evolution.\par


\subsection{Information Entropy}
Now, we will discuss the role of information entropy in measuring (de)localization~\cite{ravi}. It can be written as~\cite{katsanos, farzadian, coppola}, 
\begin{equation}
    S(t) = - \sum_{n} P_{n} \log P_{n}
\end{equation}
Where the probability of finding the wave packet in its $n^{th}$ state is denoted by $P_n = |\psi_{n}(t)|^2$ ($0 \leqslant P_{n} \leqslant 1$) and $\sum_{n} P_{n} = 1$. \\
\begin{figure}[ht!]
\centering
(a)\includegraphics[width=.43\columnwidth]{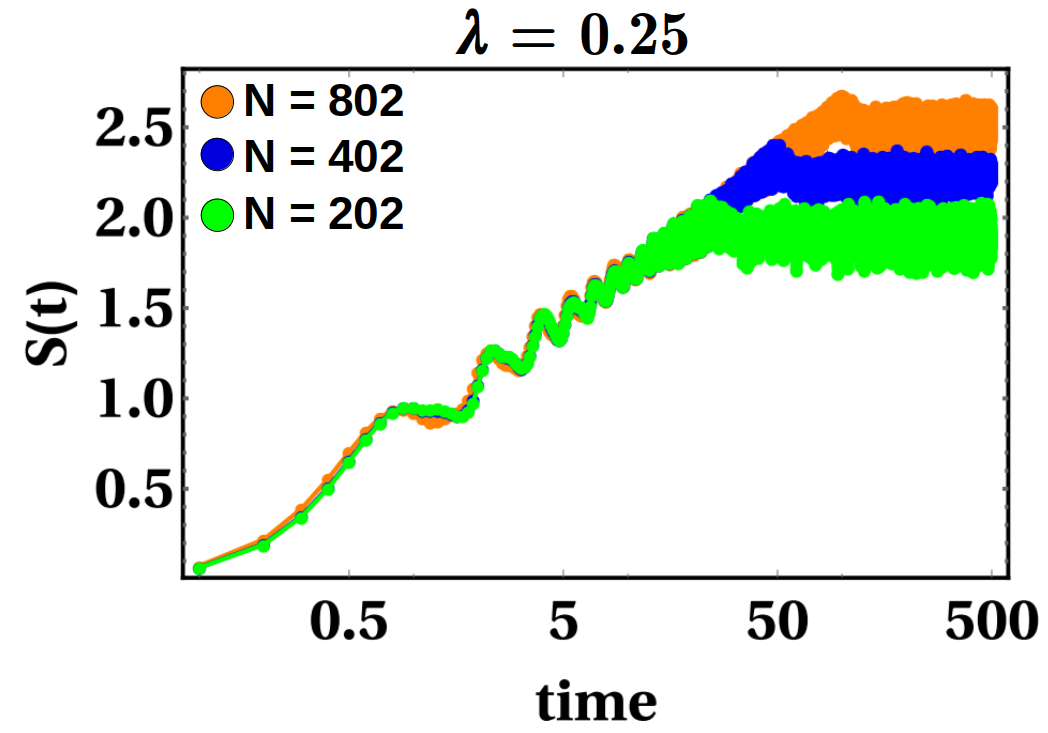}
(b)\includegraphics[width=.43\columnwidth]{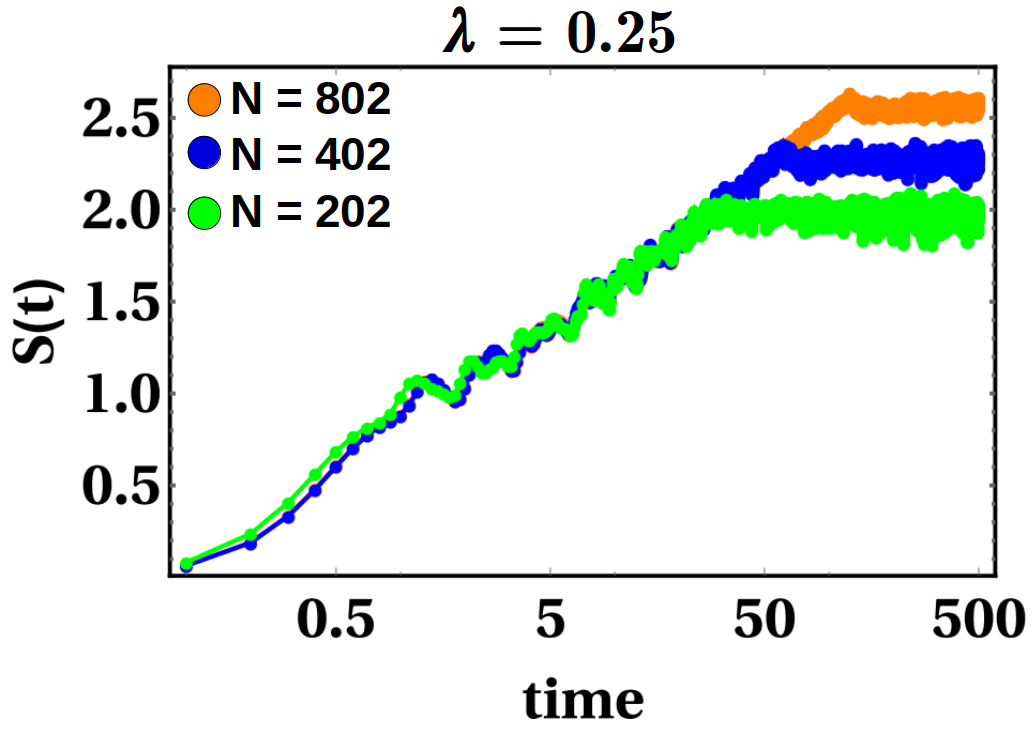}
(c)\includegraphics[width=.43\columnwidth]{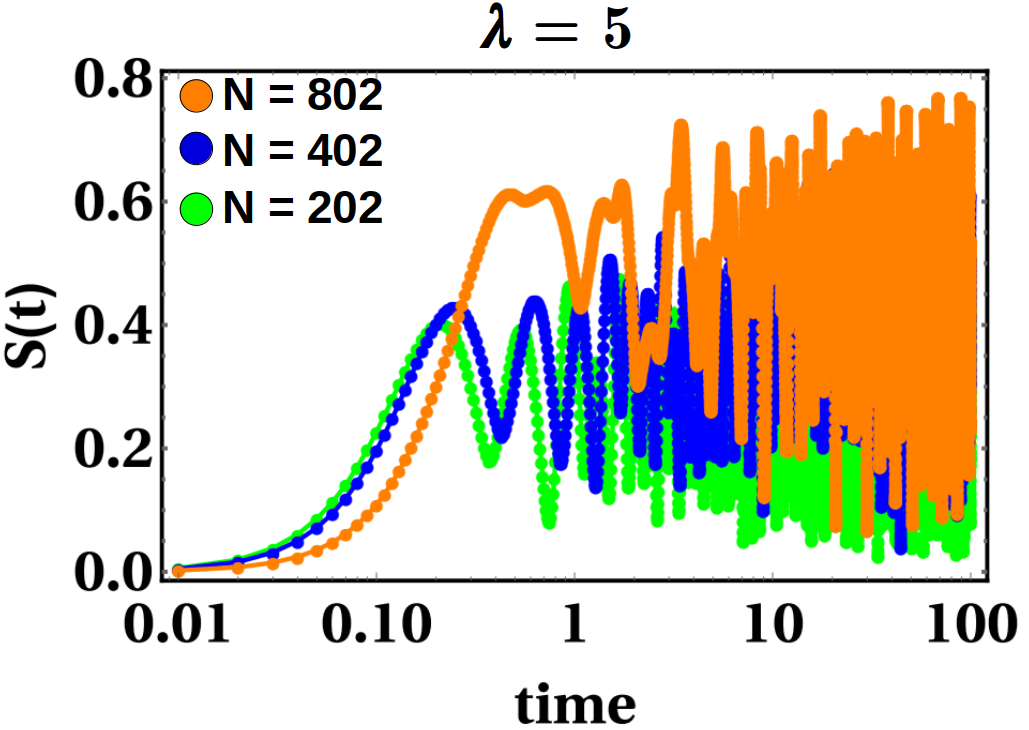}
(d)\includegraphics[width=.43\columnwidth]{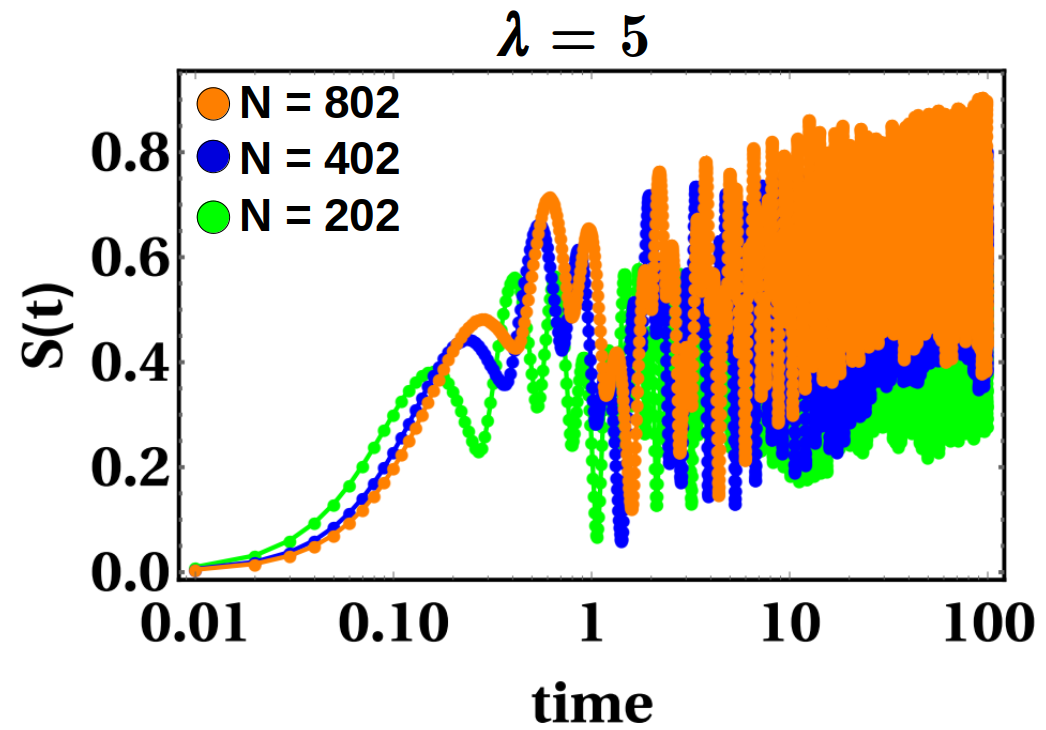}
\caption{(Color online) The information entropy (IE) is plotted against time (in a log scale) for (a,c) the ladder network and (b,d) the ultrathin graphene nanoribbon network with different system sizes, $N=202$ (green colored), $N=402$(blue colored), and $N=802$ (orange colored). Initially, we put the wavepacket at $m = 101^{th}$, $m = 201^th$, and $m = 401^{th}$ site for the system size $N=202, 402$, and $802$ respectively. The vertical (rung) hopping parameter is given by $\Gamma_{n}=y (1+ \lambda \cos( 2 \pi Q n))$, where $Q = \frac{\sqrt{5}-1}{2}$. The strength of the AAH modulation is chosen as (a,b) $\lambda = 0.25$, and (c,d) $\lambda = 5$. The onsite potential $\epsilon$ is chosen as zero. The parameter $ y $ is set as $1$. }
\label{ie}
\end{figure}
Fig.~\ref{ie} (a,b) shows the behavior of information entropy with time for the ladder and ultrathin GNR network respectively when the applied AAH modulation strength is taken as $\lambda = 0.25$. At this strength of AAH modulation, all eigenstates become delocalized and the wavepacket moves away from its initial location as time flows on. So with an increment of time, the probability of finding the wavepacket in each site becomes finite. Due to this special type of probability distribution, the information entropy monotonically increases with time~\cite{katsanos,sougata}, and saturation of IE is observed when the wavepacket spreads out all over the system. As for higher system size, the wavepacket takes more time to completely spread out, this value of saturation also increases linearly with the system size~\cite{coppola, nakagawa}. Here we calculate the IE for three different system size with $N=202$ (green colored), $N=402$(blue colored), and $N=802$ (orange colored). When $\lambda = 0.25$, (as shown in  Fig.~\ref{ie} (a,b)) both network shows a monotonically increasing IE with time and corresponding saturation also increases linearly with system size. So the behaviour of information entropy at $\lambda = 0.25$ for both networks again ensures its de-localized character.\par
The variation of IE with time for the ladder network with a higher modulation strength $\lambda = 5$ is depicted in Fig.~\ref{ie}(c). It is discussed earlier that all eigenstates are localized at this modulation strength, as a result, the wavepacket is always in its initial site of release. Now the probability of finding the wavepacket in its initial location is high (close to unity) and in all other location, it becomes zero. As a result (due to this type of probability distribution), IE has a very small time growth and then followed by a saturation which never depends on the system size~\cite{ravi, coppola,bardarson} (see Fig.~\ref{ie}(c)).\par
Fig.~\ref{ie}(d) shows the IE for the ultrathin GNR network with the applied modulation strength $\lambda = 5$. In this situation, the eigenstates contain both localized and de-localized states. Here localized part of the wavepacket remains in the initial site, whereas the de-localized part of the wavepacket spreads out with time. The amplitudes of the de-localized wavepacket are quite small. Still, these small values of the wavepacket amplitudes take a part in MSD calculation, it have no significant role in controlling the information entropy (also for return probability, temporal autocorrelation function, and time-dependent inverse participation ratio). As the localized part of the wavepacket contains higher amplitude, here also IE shows a system-independent rapid saturation with a very small time growth (see Fig.~\ref{ie}(d)).\par

\section{Conclusion}
We have investigated in detail the behaviour of eigenstates, multifractality, and quantum dynamics of a wavepacket released inside two decorated lattices under AAH modulation in their vertical arms. The ladder network shows the metal-insulator transition depending on the strength of the AAH modulation which is applied in the vertical arm. Whereas a perfect metal-to-insulator phase transition is not possible in the ultrathin GNR due to the modulation strength in-depended extendedness of the central energy region. Apart from these, both networks have a multifractal character near their transition point(critical region). The quantum dynamics of the wavepacket for these networks show completely different characters. For the ladder network, the results are as expected. From the long-time behaviour of mean square displacement, it is obvious that the system moves from ballistic motion to super-ordinary-sub-diffusion motion, and then enters its localized region as the strength of the AAH modulation is increased. The return probability, temporal autocorrelation function, and corresponding time-dependent inverse participation ratio completely support the results that come from MSD calculation. As already discussed in detail, the long-time behaviour of MSD of ultrathin GNR network always exhibits a ballistic (or very close to it) character under any strength of applied AAH modulation in the vertical arm. Due to the extendedness of the central energy region in this network, the wavepacket contains both localized and delocalized parts. With time evolution the de-localized part of the wavepacket spreads out but the localized part of the wavepacket remains in the initial site of release. MSD is mainly controlled by the delocalized part of the wavepacket as the initial site's amplitude never affects the MSD calculation. In general, if a system shows ballistic motion it has no return probability as the wavepacket moves throughout the system and the corresponding inverse participation ratio goes to zero as time flows on. But in the ultrathin GNR case, the localized part of the wavepacket is still in the initial location after a long time, as a result, the system has a finite return probability when the system enters its mixed region (both delocalized and localized states) and corresponding IPR never decay to zero with increment of time. So we can conclude that the quantum dynamics of the mixed states have a special character, MSD is controlled by the delocalized part of the wavepacket whereas the localized part of the wavepacket plays the main role in determining the return probability, temporal auto-correlation function, information entropy, and time-dependent inverse participation ratio. 

\section{Acknowledgements}
I want to express my sincere gratitude to my supervisor Prof. Arunava Chakrabarti for his guidance, invaluable advice, and continuous support while carrying on the present research work. I am thankful to the Government of West Bengal for the SVMCM Scholarship (WBP221657867058).

\appendix


\section{}
\begin{figure}[ht!]
\centering
(a)\includegraphics[width=.75\columnwidth]{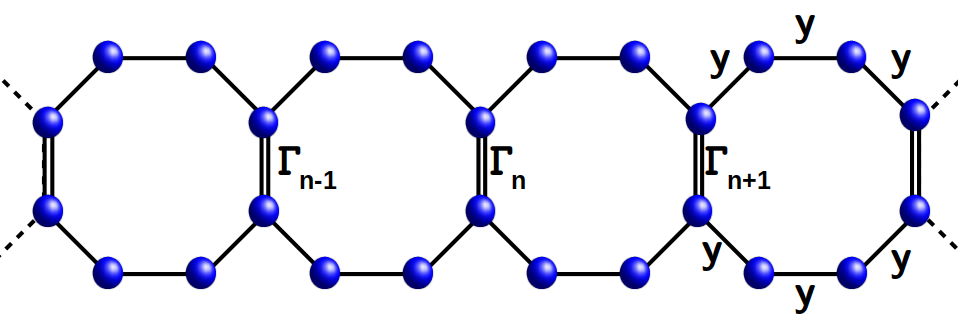}
(b)\includegraphics[width=.5\columnwidth]{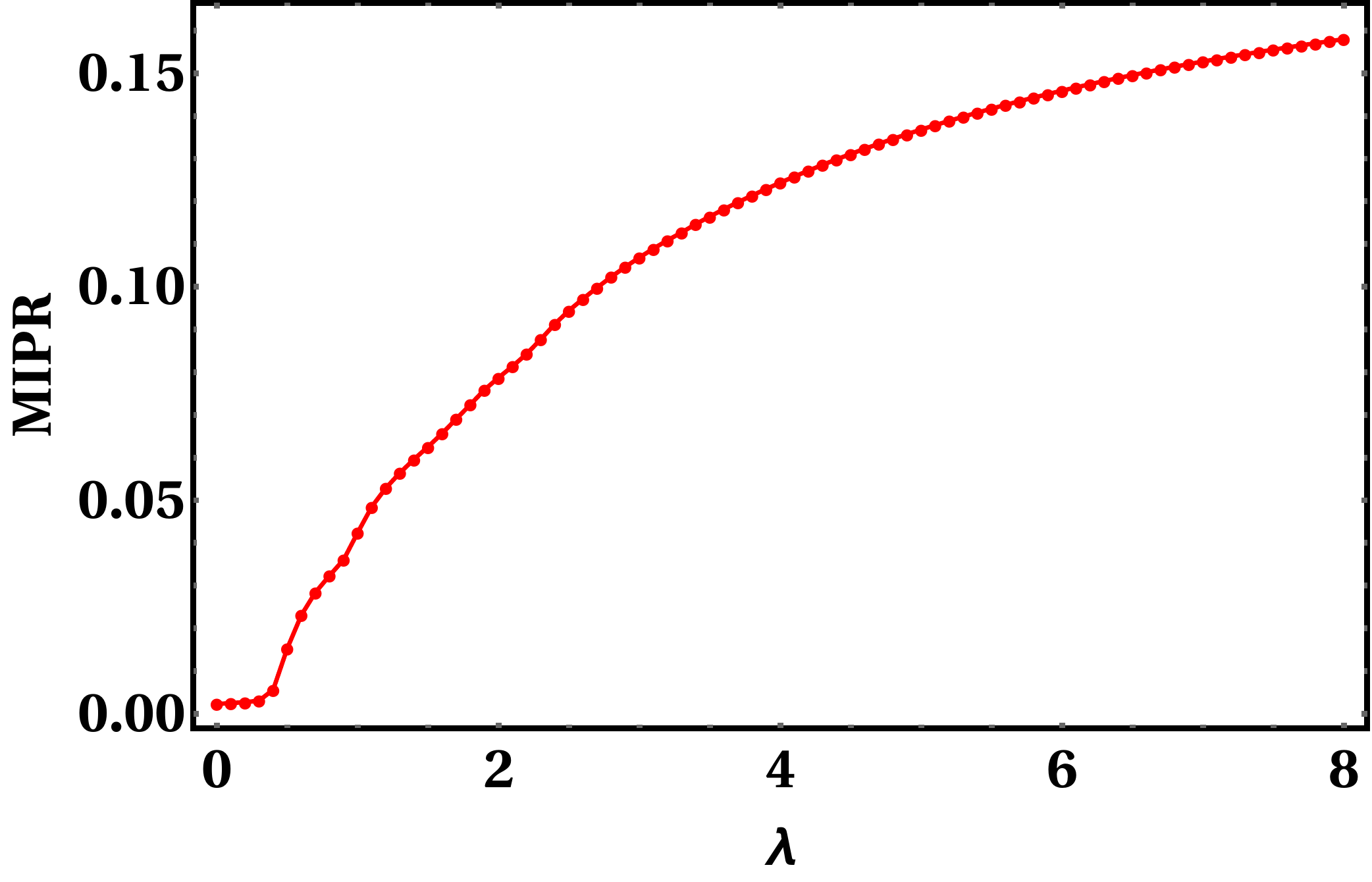}
(c)\includegraphics[width=.42\columnwidth]{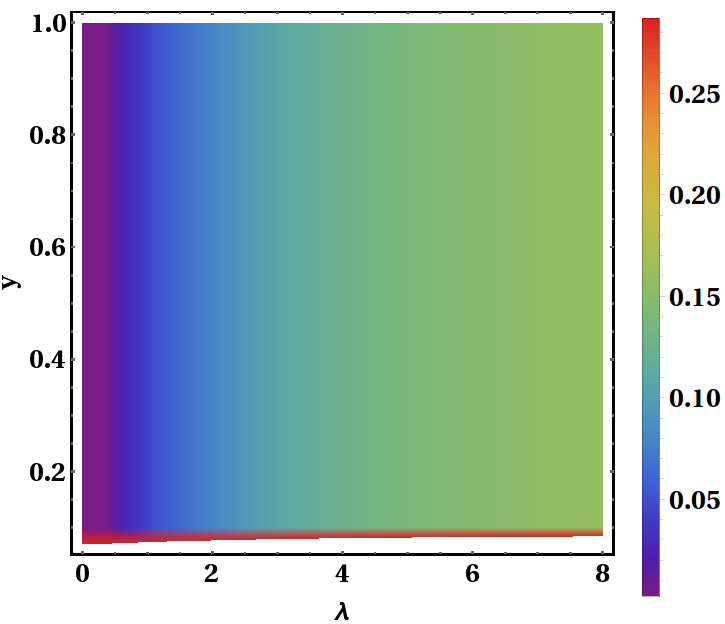}
(d)\includegraphics[width=.42\columnwidth]{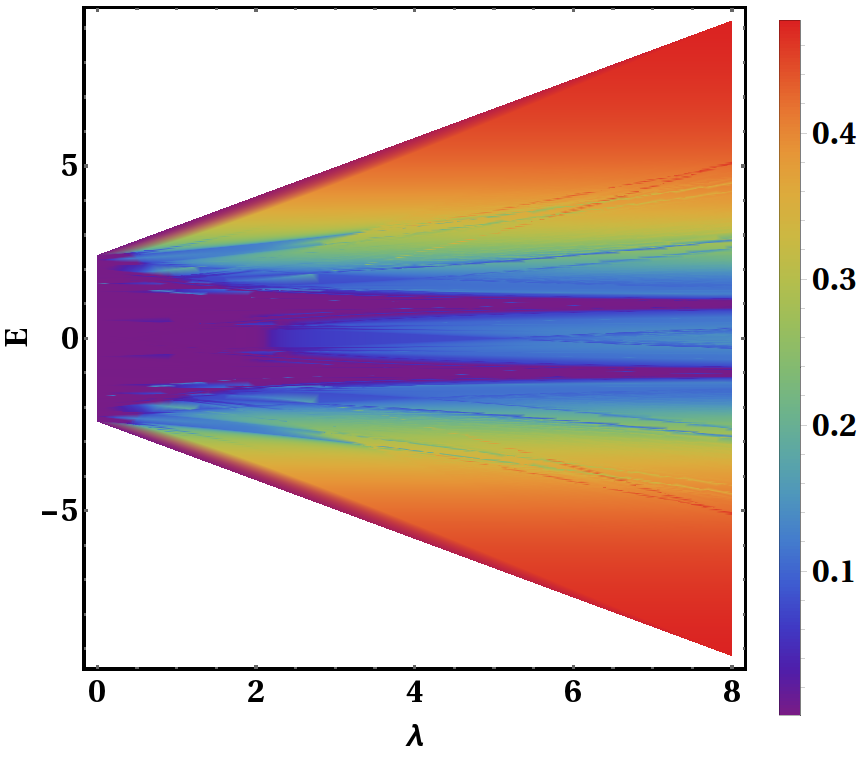}
\caption{(Color online) (a) Another variant of the ladder network with Aubry modulation in its rung hopping part. (b) Mean Inverse Participation Ratio (MIPR) is plotted against $\lambda$. The rung hopping parameter is given by $\Gamma_{n}= y (1+ \lambda \cos( 2 \pi Q n))$, where $Q = \frac{\sqrt{5}-1}{2}$. The onsite potential $\epsilon$ is chosen as zero. The parameter $ y $ is set as $1$. (c) Density plot of MIPR within the parameter space $y$ and $\lambda$. (d) Density plot of IPR with energy ($E$) and $\lambda$. }
\label {octagon}
\end{figure}

Here we discuss another decorated lattice model, another variant of the ladder network (as shown in Fig.~\ref{octagon}(a)) with an applied AAH modulation in the vertical arm (same as previously selected lattice models).

Fig.~\ref{octagon}(b) shows the variation of mean inverse participation ratio with the strength of the AAH modulation $\lambda$ and the corresponding density plot of MIPR is depicted in Fig.~\ref{octagon}(c). As discussed earlier the central energy band of the ultrathin GNR is always delocalized irrespective of modulation strength $\lambda$, similar behaviour of eigenstates is followed in this ladder variant Fig.~\ref{octagon}(d). It also shows multifractality near the transition point and quantum dynamics are completely similar to the ultrathin GNR network.

\end{document}